\documentclass[10pt]{article}
\usepackage{amssymb}
\usepackage{amsmath}
\usepackage{graphicx}
\usepackage{acronym}
\usepackage{amsfonts}

\newtheorem{theorem}{Theorem}[section]
\newtheorem{corollary}{Corollary}[section]

\newtheorem{lemma}{Lemma}[section]
\newtheorem{definition}{Definition}

\newtheorem{Proposition}{Proposition}[section]
\newtheorem{Example}{Example}[section]
\newtheorem{Assumption}{Assumption}[section]
\newtheorem{Algorithm}{Algorithm}[section]
\newtheorem{Remark}{Remark}[section]
\newtheorem{remark}{remark}[section]

\def\be{\begin{Example}}
\def\ee{\end{Example}}
\def\bt{\begin{theorem}}
\def\et{\end{theorem}\vskip 6pt}
\def\bl{\begin{lemma}}
\def\el{\end{lemma}}
\def\ep{\end{Proposition}}
\def\bp{\begin{Proposition}}
\def\bd{\begin{definition}}
\def\ed{\end{definition}}
\def\ba{\begin{Algorithm}}
\def\ea{\end{Algorithm}}
\def\bs{\begin{Assumption}}
\def\es{\end{Assumption}}
\def\br{\begin{Remark}}
\def\er{\end{Remark}}

\input epsf
\begin{document}

\title{Single-photon quantum filtering with multiple measurements}

\author{Zhiyuan Dong\thanks{Zhiyuan Dong is with the Department of Applied Mathematics, the Hong Kong
Polytechnic University, Hong Kong, China (email: zhiyuan.dong@connect.polyu.hk).} \and  Guofeng Zhang\thanks{Guofeng Zhang is with the Department of Applied Mathematics, the Hong Kong
Polytechnic University, Hong Kong, China (email: guofeng.zhang@polyu.edu.hk).}  \and Nina H. Amini\thanks{Nina H. Amini is with CNRS, Laboratoire des signaux et syst\`{e}mes (L2S), CentraleSup\'{e}lec, 3 rue Joliot Curie, 91192 Gif-Sur-Yvette, France  (email: nina.amini@lss.supelec.fr).}}


\date{\today}

\maketitle

\abstract{The single-photon quantum filtering problems have been investigated recently with applications in quantum computing. In practice, the detector responds with a quantum efficiency of less than unity since there exists some mode mismatch between the detector and the system, and the single-photon signal may be corrupted by quantum white noise. Consequently, quantum filters based on multiple measurements are designed in this paper to improve the estimation performance. More specifically, the filtering equations for a two-level quantum system driven by a single-photon input state and under multiple measurements are presented in this paper. Four scenarios, 1) two diffusive measurements with Q-P quadrature form, 2)  two diffusive measurements with Q-Q quadrature form, 3) diffusive plus Poissonian measurements, and 4) two Poissonian measurements, are considered. It is natural to compare the filtering results, i.e., measuring single channel or both channels, which one is better? By the simulation where we use a single photon to excite an atom, it seems that multiple measurements enable us to excite the atom with higher probability than only measuring single channel. In addition, measurement back-action phenomenon is revealed by the simulation results.}

Keywords: quantum filtering; single-photon state; homodyne detection; photon-counting; quantum trajectories

\medskip

\section{Introduction}

Over the past few decades, quantum filtering has drawn researchers' lots of attention and been rapidly developed \cite{bouten2007introduction,chia2011quantum,RNC:RNC3530}. Its modern form and foundational framework were firstly studied by Belavkin in \cite{belavkin1989nondemolition,belavkin1995quantum}. Particularly in quantum optics, quantum filtering is known as master equation and stochastic master equation. The latter represents the stochastic evolution of the conditional density operator when the system interacts with the field. The quantum trajectory theory, which can describe this stochastic process, is developed by Carmichael in \cite{carmichael2009open} and has been widely applied in quantum filtering and quantum control \cite{GJN13,GJN12,GOUGH12QUANTUM,GZ15,SONG13MULTI,ZJ11,RNC:RNC3577}.

The framework of quantum filtering for system driven by Gaussian input fields, such as coherent state, squeezed state, thermal state and vacuum state, have been well treated in a series of articles \cite{dum1992monte,gardiner2004quantum,Hendra14QUANTUM,wiseman2009quantum}. A two-level atom driven by a single-photon state is considered in \cite{Gheri1998photon}, and master equations have been derived in detail to illustrate the formalism presented is applicable to $N$-photon wave packets. Filtering equations for systems driven by single-photon states or superposition of coherent states have been derived in \cite{GJN13,GJN12,GOUGH12QUANTUM,dong2015quantum}. Particularly, an ancilla system is introduced to model the effect of the single-photon input state on the system. Non-classical states, single-photon states and coherent states, have been considered in \cite{GOUGH12QUANTUM}. The stochastic master equations for the whole extended system with homodyne detection and photon-counting measurements are given respectively. A non-Markovian embedding method was conducted in \cite{GJN13,GJN12}, the ancilla, system and field are supposed to be in a superposition state initially. The proposed framework with this embedding method could derive the filtering equations for system driven by non-classical fields efficiently.

In quantum optics experiments, there may exist some limitations due to the impure input state and the imperfect measurements. A set of experimental imperfections in the input, the photon subtraction, and the detector are taken into account in \cite{SKH2013Limitations}. The effects of various experimental parameters on the Schr\"{o}dinger kitten state generation have been discussed in terms of non-Gaussian property witness and the origin of Wigner function. A finite dimensional Markov system with quantum non-demolition measurements is considered in \cite{amini2013feedback}. Quantum filters and robustness property for both perfect and imperfect measurements have been analyzed, the convergence of the controlled system is ensured when imperfect measurements corrupted by random errors. Both continuous-time and discrete-time quantum filters with incompleteness and errors in measurements are presented in \cite{amini2014stability}. Particularly, stochastic master equations (SMEs) for quantum system driven by Poisson and Wiener processes are discussed and the detection errors are modeled by a random matrix. An experimental implementation of quantum feedback set-up has been conducted by using the photon box and closed-loop simulations are also presented in \cite{Sayrin2011}.

Recently, based on the general framework for single-photon filtering \cite{GOUGH12QUANTUM}, the stochastic master equations for quantum systems driven by a single-photon input state which is contaminated by quantum vacuum noise have been presented in \cite{dong2015quantum}. The composite state is prepared as $|1_\xi\rangle\otimes|0\rangle$, where $|1_\xi\rangle$ means the single-photon state and $|0\rangle$ is the vacuum noise. By the input-output formalism  \cite{GC85,gardiner2004quantum,carmichael2009open,GOUGH09SERIES}, the output field state would be a superposition state $|\psi_{\mathrm{out}}\rangle=s_{11}|1_\eta\rangle\otimes|0\rangle+s_{21}|0\rangle\otimes|1_\eta\rangle$, where $\eta$ is the output pulse shape, and  $s_{11}$, $s_{21}$ are parameters of the beam splitter, Fig. \ref{fig_3}. Thus, the filter may be more efficient if both output channels are measured. Quantum filters with diffusive plus Poissonian measurements and two diffusive measurements have been designed in \cite{dong2015quantum} to improve atom excitation efficiency. Moreover, when there is no vacuum noise and mode mismatch, the scenario will be reduced to the ideal case which has been discussed in \cite{GOUGH12QUANTUM}. In this paper, we extend the single-photon filtering results in \cite{dong2015quantum}. More specifically, instead of the simple beam splitter used in  \cite{dong2015quantum} which is given by Eq. (\ref{sb2}) in the present paper, a more general form of the beam splitter is used, see Eqs. \eqref{S_b} and \eqref{B} for more details. As a result, the stochastic master equations contain more parameters and are more general than those in \cite{dong2015quantum}. More importantly,  a comprehensive study of numerical studies is presented in this paper. In particular, we consider all combinations of measurement settings, namely, two diffusive measurements with Q-Q and Q-P combinations (subsection \ref{twohomodyne}), diffusive plus Poissonian measurements (subsection \ref{homodyne+photon-counting}),  and two Poissonian measurements (subsection \ref{twophoton-counting}). For example, in Fig. \ref{fig_5}, when there is no vacuum noise and mode mismatch, we recover the results in \cite{GOUGH12QUANTUM}. In Fig. \ref{fig_6}, we consider the case of two homodyne detection measurements as the case discussed in \cite{dong2015quantum}. However, we go beyond the simulations done in \cite{dong2015quantum} by considering more beam splitter parameters. In addition, in Fig. \ref{fig_7} and Fig. \ref{fig_8} we consider the homodyne detection plus photon-counting measurements and two photon-counting measurements respectively, which have not been discussed in the literature. By comparing  Fig. \ref{fig_7} and Fig. \ref{fig_6} (a), (c), and (e), it can be seen that for some parameters, the performance of the mixed measurements is better than that of the single homodyne measurement, but this is not true in general. Finally, from Fig. \ref{fig_8} one can see that photon-counting measurement reflects the particle nature of photons.

This paper is structured as follows. In section \ref{preliminary}, we recall some basic theory such as open quantum systems, series products, quantum filtering and continuous-mode single-photon state. Quantum filters with multiple measurements have been discussed in section \ref{quantumfilter}. Firstly, filtering equations for two homodyne detectors with different quadrature forms, i.e., Q-P and Q-Q measurements, have been presented explicitly in subsection \ref{twohomodyne}. Then in subsection \ref{homodyne+photon-counting}, quantum filters are given in the case of joint homodyne detection and photon-counting measurements. Two Poissonian measurements is also considered in subsection \ref{twophoton-counting}. Simulation results in subsection \ref{simulation} verify the effectiveness of the proposed framework. We conclude this paper in Section \ref{conclusions}.

\emph{Notation}. Let $i$ be the imaginary unit and $|0\rangle$ be the vacuum state of the free field. Serif symbols are used for Hilbert spaces, e.g. \textsf{H}. The Hilbert space adjoint or complex conjugate is indicated by $\ast$. The complex conjugate transpose will be denoted by $\dag$, i.e. $X^\dag=(X^\ast)^T$. We will use $\ast$ and $\dag$ interchangeably for single-element operators. The inner product of $X$ and $Y$ in Hilbert space is given by $\langle X,Y\rangle$. $[A,B]=AB-BA$ denotes the commutator between operators $A$ and $B$. With the triple language description of a system, $G=(S,L,H)$, the associated superoperators are defined as
\begin{equation}\nonumber\begin{aligned}
{\rm Lindbladian}:~&\mathcal{L}_GX\equiv-i[X,H]+\mathcal{D}_LX,\\
{\rm Liouvillian}:~&\mathcal{L}^\star_G\rho\equiv-i[H,\rho]+\mathcal{D}^\star_L\rho,
\end{aligned}\end{equation}
where $\mathcal{D}_AB\equiv A^\dag BA-\frac{1}{2}(A^\dag AB+BA^\dag A)$ and $\mathcal{D}^\star_AB\equiv ABA^\dag-\frac{1}{2}(A^\dag AB+BA^\dag A)$. Obviously, we have $\mathrm{Tr}[\rho\mathcal{L}_GX]=\mathrm{Tr}[X\mathcal{L}^\star_G\rho]$ for traceclass $\rho$ and bounded $X$. Finally, $\otimes$ denotes the tensor product.

\section{Preliminary}\label{preliminary}

\subsection{Open quantum systems}

The system model we discuss is a two-level quantum system driven by a single-photon input field. Here, we will describe the system by using the $(S,L,H)$ formalism \cite{GOUGH09SERIES,ZJ12}. The scattering operator $S$ is unitary, which satisfies $S^{\dag}S=SS^{\dag}=I$. The coupling between system and field is described by the operator $L$ and the self-adjoint operator $H$ is the initial Hamiltonian of the system.

The input field is represented by annihilation operator $b(t)$ and creation operator $b^\dag(t)$ on the Fock space $\textsf{H}_F$, which satisfy $[b(t),b^\dag(s)]=\delta(t-s)$. The integrated annihilation and creation operators, together with the gauge process are given by
\begin{equation}\nonumber\begin{aligned}
B(t)=\int^t_0b(s)ds,~~B^\dag(t)=\int^t_0b^\dag(s)ds,~~\Lambda(t)&=\int^t_0b^\dag(s)b(s)ds.
\end{aligned}\end{equation}

In this paper, we assume that these quantum stochastic processes are canonical, that is, their products satisfy the following It\={o} table
\begin{equation}\begin{aligned}
\begin{array}{c|cccc}
  \times & dt & dB & d\Lambda & dB^\dag \\ \hline
  dt & 0 & 0 & 0 & 0 \\
  dB & 0 & 0 & dB & dt \\
  d\Lambda & 0 & 0 & d\Lambda & dB^\dag \\
  dB^\dag & 0 & 0 & 0 & 0
\end{array}.
\end{aligned}\end{equation}
In fact, the single-photon input field and vacuum input field studied in this paper satisfy the above It\={o} table.

The dynamical evolution can be described by a unitary operator $U(t)$ on the tensor product Hilbert space $\textsf{H}_S\otimes\textsf{H}_F$ which is given by the following quantum stochastic differential equation (QSDE)
\begin{equation}\begin{aligned}
dU(t)=\bigg\{{\rm Tr}[(S-I)d\Lambda(t)]+LdB^\dag(t)-L^{\dag}SdB(t)-\left(\frac{1}{2}L^{\dag}L+iH\right)dt\bigg\}U(t),
\end{aligned}\end{equation}
where $U(0)=I$.

In Heisenberg picture, the system operator $X$ is given by a joint operator $j_t(X)=U^{\dag}(t)(X\otimes I_{\mathrm{field}})U(t)$ on $\textsf{H}_S\otimes\textsf{H}_F$. By the quantum It\={o} product rule and table, the temporal evolution of $j_t(X)\equiv X(t)$ is derived as
\begin{equation}\begin{aligned}
dj_t(X)=j_t(\mathcal{L}_GX)dt+j_t([L^\dag,X]S)dB(t)+j_t(S^\dag[X,L])dB^{\dag}(t)+{\rm Tr}[j_t(S^{\dag}XS-X)d\Lambda(t)].
\end{aligned}\end{equation}
The output fields are defined by
\begin{equation}\nonumber\begin{aligned}
B_{\mathrm{out}}(t)&=U^\dag(t)(I_{\mathrm{system}}\otimes B(t))U(t),\\
\Lambda_{\mathrm{out}}(t)&=U^\dag(t)(I_{\mathrm{system}}\otimes\Lambda(t))U(t),
\end{aligned}\end{equation}
and by It$\bar{\mathrm{o}}$ calculus, we can find the following evolution
\begin{equation}\nonumber\begin{aligned}
dB_{\mathrm{out}}(t)=&S(t)dB(t)+L(t)dt,\\
d\Lambda_{\mathrm{out}}(t)=&S^\ast(t)d\Lambda(t)S^T(t)+S^\ast(t)dB^\ast(t)L^T(t)+L^\ast(t)dB^T(t)S^T(t)+L^\ast(t)L^T(t)dt.
\end{aligned}\end{equation}

\subsection{The concatenation and series products}

\subsubsection{Concatenation product}

\begin{figure}
\centering
\includegraphics[width=0.5\textwidth]{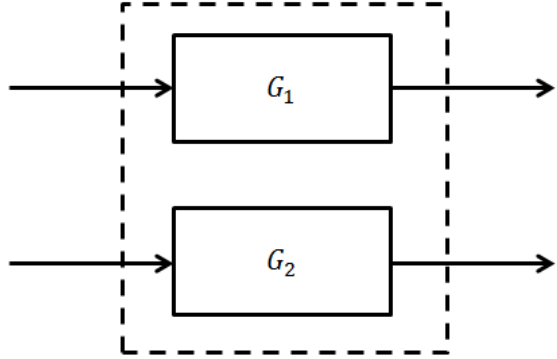}
\caption{Concatenation product.}\label{Concatenation product}
\end{figure}

Given two systems $G_1=(S_1,L_1,H_1)$ and $G_2=(S_2,L_2,H_2)$, see Fig.~\ref{Concatenation product}, we define the concatenation product \cite{GOUGH09SERIES} to be the system $G_1\boxplus G_2$ by
\begin{equation}\nonumber\begin{aligned}
G_1\boxplus G_2=\left(\left[
                         \begin{array}{cc}
                           S_1 & 0 \\
                           0 & S_2 \\
                         \end{array}
                       \right]
,\left[
                          \begin{array}{c}
                            L_1 \\
                            L_2 \\
                          \end{array}
                        \right]
,H_1+H_2\right).
\end{aligned}\end{equation}

\subsubsection{Series product}

\begin{figure}
\centering
\includegraphics[width=0.5\textwidth]{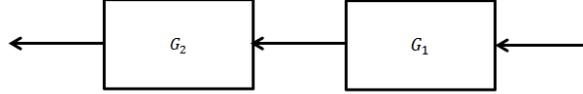}
\caption{Series product.}
\label{Series product}
\end{figure}

Given two systems $G_1=(S_1,L_1,H_1)$ and $G_2=(S_2,L_2,H_2)$ with the same number of field channels, see Fig.~\ref{Series product}, we define the series product \cite{GOUGH09SERIES} to be the system $G_2\vartriangleleft G_1$ by
\begin{equation}\nonumber\begin{aligned}
G_2\vartriangleleft G_1=\left(S_2S_1,L_2+S_2L_1,H_1+H_2+\mathrm{Im}\{L^\dag_2S_2L_1\}\right).
\end{aligned}\end{equation}

\subsection{Quantum filtering}

Homodyne and photon-counting detections are the most commonly used measurement methods in quantum optics \cite{gardiner2004quantum, wiseman2009quantum,bouten2007introduction,SONG13MULTI}. By using homodyne detection, the measurement is given by quadrature phase
\begin{equation}\nonumber\begin{aligned}
Y(t)=U^\dag(t)(I_{\mathrm{system}}\otimes(B(t)+B^\ast(t)))U(t) = B_{\rm out}(t)+ B_{\rm out}^\ast(t),
\end{aligned}\end{equation}
while in the photon-counting case the measurement outcomes are photon numbers
\begin{equation}\nonumber\begin{aligned}
Y(t)=U^\dag(t)(I_{\mathrm{system}}\otimes\Lambda(t))U(t)=\Lambda_{\rm out}(t).
\end{aligned}\end{equation}
Both of the measurements satisfy the following self-nondemolition relations, \cite[Section 5.2]{bouten2007introduction},
\begin{equation}\nonumber\begin{aligned}
\left[Y(s),Y(t)\right]=0,~~0\leq s\leq t.
\end{aligned}\end{equation}
The quantum conditional expectation is given by
\begin{equation}\nonumber\begin{aligned}
\hat{X}(t)=\pi_t(X)=\mathbb{E}[j_t(X)|\mathcal{Y}_t],
\end{aligned}\end{equation}
where $\mathcal{Y}_t$ is generated by $\{Y(s):0\leq s\leq t\}$. Generally speaking, the quantum filtering problem is about minimizing the least mean-squares estimate $\mathbb{E}[\{\hat{X}(t)-j_t(X)\}^2]$ of system observables $j_t(X)$ based on the past measurement information $\mathcal{Y}_t$. Notice that the quantum conditional expectation is well-defined since it satisfies the non-demolition property $[j_t(X),Y(s)]=0$ for all $0\leq s\leq t$.

\subsection{Continuous-mode single-photon state}\label{singlephoton}

The creation operator for a photon with wave packet $\xi(t)$ in the time domain is defined as
\begin{equation}\nonumber\begin{aligned}
B^\ast(\xi)=\int^{\infty}_0\xi(t)b^\ast(t)dt,
\end{aligned}\end{equation}
with the normalization condition $\int^\infty_0|\xi(t)|^2dt=1$. Then the single-photon state \cite{loudon2000quantum,rohde2005frequency,milburn2008coherent} is given by
\begin{equation}\nonumber\begin{aligned}
|1_{\xi}\rangle=B^\ast(\xi)|0\rangle.
\end{aligned}\end{equation}
Similarly, we can define the single-photon state in the frequency domain
\begin{equation}\nonumber\begin{aligned}
|1_{\xi}\rangle=\int^{\infty}_{-\infty}\hat{\xi}(\omega)\hat{b}^\ast(\omega)d\omega|0\rangle,
\end{aligned}\end{equation}
where $\hat{\xi}(\omega)$ is the Fourier transform of $\xi(t)$.

The original difference of the master equations and stochastic master equations between single-photon input and vacuum input is given by the following identities
\begin{equation}\nonumber\begin{aligned}
dB(t)|1_\xi\rangle=\xi(t)dt|0\rangle, \ \ \ dB(t)|0\rangle = 0, \\
d\Lambda(t)|1_\xi\rangle=\xi(t)dB^\dag(t)|0\rangle, \ \ \ d\Lambda(t)|0\rangle = 0.
\end{aligned}\end{equation}

\section{Quantum Filter for Multiple Measurements}\label{quantumfilter}

\begin{figure}
\centering
\includegraphics[width=0.7\textwidth]{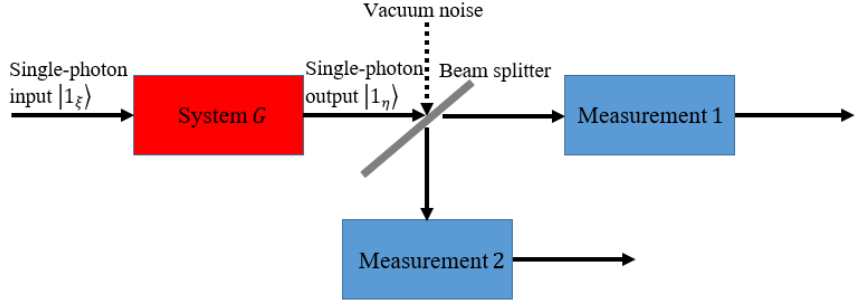}
\caption{\label{fig_3}(Color online) Schematic representation of multiple measurements at the outputs in a quantum system.}
\end{figure}

In this section, we mainly consider quantum filtering for a two-level system $G=(S,L,H)$ interacting with a single photon, see Fig. \ref{fig_3}. Due to the impure input state and the imperfect measurements, vacuum noise is considered as dashed curve in Fig. \ref{fig_3}. Based on multiple measurements, a beam splitter with general form is used to design quantum filters. Particularly, the beam splitter parameters can be chosen to compare the weights of the two measurements in Fig. \ref{fig_3}. The extended system is briefly reviewed and the relation between expectation for system and for extended system is discussed in subsection~\ref{extended}, the quantum filter for system driven by vacuum state is introduced in subsection~\ref{vacuuminput}. For the single-photon input field, we present the filtering equations in subsection~\ref{twohomodyne}, \ref{homodyne+photon-counting} and \ref{twophoton-counting}. An illustrating example is given in subsection \ref{simulation}.

\subsection{The extended system}\label{extended}

In this subsection, we use the method of single-photon generation in \cite{GOUGH12QUANTUM,GZ15} to generate a single photon from a vacuum field. The quantum signal generating filter $M=(S_M,L_M,H_M)$, which is usually called ancilla, is cascaded with our quantum two-level system $G$. Then the triple parameters for the extended system $G_T=G\vartriangleleft M$ can be derived by the series product. Since master equations and filtering equations for quantum system driven by vacuum input state have already been studied, e.g. \cite{GC85,dum1992monte,gardiner2004quantum,GOUGH12QUANTUM}, we can present the quantum filters for the whole system directly. The triple parameters of signal generating filter is given by
\begin{equation}\nonumber\begin{aligned}
(S_M,L_M,H_M)=(I,\lambda(t)\sigma_-,0),
\end{aligned}\end{equation}
where $\sigma_-$ is the lowering operator from the upper state $|\uparrow\rangle$ to the ground state $|\downarrow\rangle$. The ancilla, which is a two-level atom, is initialized in the upper state $|\uparrow\rangle$. Driven by the vacuum field, it will  decay into its ground state and generate a single photon. The ancilla will generate the desired single-photon state $|1_\xi\rangle$ if the coupling strength $\lambda(t)$ is chosen to be
\begin{equation}\nonumber\begin{aligned}
\lambda(t)=\frac{\xi(t)}{\sqrt{w(t)}},
\end{aligned}\end{equation}
where $w(t)=\int^{\infty}_t|\xi(s)|^2ds$, $t\geq0$.

By means of series product, we have the extended system $G_T$
\begin{equation}\nonumber\begin{aligned}
G_T=G\vartriangleleft M=\left(S,L+\lambda(t)S\sigma_-,H+\lambda(t)\mathrm{Im}\{L^{\dag}S\sigma_-\}\right).
\end{aligned}\end{equation}

Let $\tilde{U}(t)$ be the unitary operator for the joint ancilla-system-field system. The following equality can be shown (see \cite{GOUGH12QUANTUM} for more details)
\begin{equation}\nonumber\begin{aligned}
\mathbb{E}_{\eta\xi}[X(t)]=\mathbb{E}_{\uparrow\eta0}[\tilde{U}^\dag(t)(I\otimes X\otimes I)\tilde{U}(t)],
\end{aligned}\end{equation}
with initial state $|\uparrow\rangle\otimes|\eta\rangle\otimes|0\rangle$ for an arbitrary operator $X(t)$ of the system $G$.

\subsection{Quantum filter for multiple measurements driven by vacuum input}\label{vacuuminput}

The result of multiple measurements with vacuum state input is needed to present the quantum filter for system driven by a single-photon input state.
\begin{lemma}\label{peterlemma} \cite{emzir2015quantum}
Let $\{Y_{i,t},i=1,2,\ldots,N\}$ be a set of $N$ compatible measurement outputs for a quantum system $G$. With vacuum initial state, the corresponding joint measurement quantum filter is given by
\begin{equation}\begin{aligned}
d\hat{X}=\pi_t[\mathcal{L}_G(X_t)]dt+\displaystyle{\sum^N_{i=1}}\beta_{i,t}dW_{i,t},
\end{aligned}\end{equation}
where $dW_{i,t}=dY_{i,t}-\pi_t(dY_{i,t})$ is a martingale process for each measurement output and $\beta_{i,t}$ is the corresponding gain given by
\begin{equation}\nonumber\begin{aligned}
\zeta^T&=\pi_t(X_tdY^T_t)-\pi_t(X_t)\pi_t(dY^T_t)+\pi_t\left([L^\dag_t,X_t]S_tdBdY^T_t\right),\\
\Sigma&=\pi_t(dY_tdY^T_t), ~~ \beta=\Sigma^{-1}\zeta,
\end{aligned}\end{equation}
where $\Sigma$ is assumed to be non-singular.
\end{lemma}

\begin{Remark}
A general measurement equation, which is a function of annihilation, creation and conservation processes in the output field, is defined as \cite{emzir2015quantum}
\begin{equation}\label{fg}\begin{aligned}
dY(t)=F^\ast dB^\ast_{\mathrm{out}}(t)+FdB_{\mathrm{out}}(t)+G\mathrm{diag}(d\Lambda_{\mathrm{out}}(t)).
\end{aligned}\end{equation}
Particularly, a combination of homodyne detection and photon-counting measurement is given by
\begin{equation}\nonumber\begin{aligned}
F=\left[
    \begin{array}{cc}
      1 & 0 \\
      0 & 0 \\
    \end{array}
  \right], G=\left[
               \begin{array}{cc}
                 0 & 0 \\
                 0 & 1 \\
               \end{array}
             \right].
\end{aligned}\end{equation}
\end{Remark}

\subsection{Quantum filter for two homodyne detection measurements}\label{twohomodyne}

Assume that the system is in the initial state $\rho_0=|\eta\rangle\langle\eta|$, we define the expectation
$$\mathbb{E}_{mn}[j_t(X)]=\langle\eta\psi_m|X(t)|\eta\psi_n\rangle,~~m,n=0,1,$$
where $\psi_i=\left\{\begin{array}{ll}
                      |0\rangle, & i=0; \\
                      |1_{\xi}\rangle, & i=1.
                    \end{array}
\right.$
and the conditional expectation
\begin{equation}\nonumber\begin{aligned}
\pi^{11}_t(X)=\mathbb{E}_{\eta\xi}[X(t)|Y(s),0\leq s\leq t],
\end{aligned}\end{equation}
then the quantum filter for the extended system $G_T$ driven by the vacuum state is given by
\begin{equation}\nonumber\begin{aligned}
\tilde{\pi}_t(A\otimes X)=\mathbb{E}_{\uparrow\eta0}[\tilde{U}^\dag(t)(A\otimes X)\tilde{U}(t)|I\otimes Y(s),0\leq s\leq t],
\end{aligned}\end{equation}
where $X$ is the system operator and $A$ is an operator of the ancilla.

\begin{figure}
\centering
\includegraphics[width=0.7\textwidth]{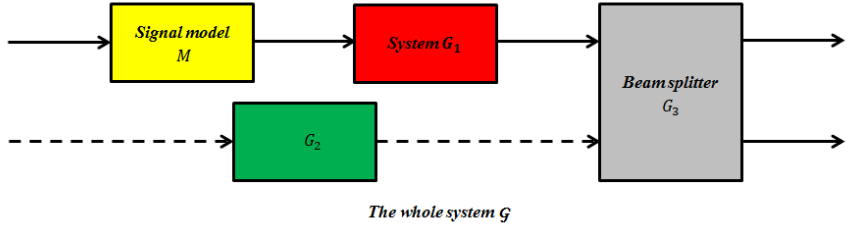}
\caption{\label{fig_4}(Color online) Quantum system depiction.}
\end{figure}

The whole system $\mathcal{G}$ with the measurements in Fig.~\ref{fig_3} can be depicted as shown in Fig.~\ref{fig_4}. $G_1=(S,L,H)$ is the original system $G$, which has been connected with a signal model (ancilla) $M=(I,L_M,0)$, where $L_M=\lambda(t)\sigma_-$. By introducing a second open quantum system $G_2=(1,0,0)$, we concatenate the vacuum noise into our system. In this paper, we consider a more general case, i.e., the last open quantum system is a beam splitter $G_3=(S_b,0,0)$, where
\begin{equation} \label{S_b}
S_b=\left[
       \begin{array}{cc}
         s_{11} & s_{12} \\
         s_{21} & s_{22} \\
       \end{array}
     \right].
\end{equation}
For example, the following beam splitter given in \cite[Eq. (4.3)]{leo03}
\begin{equation} \label{B}
B=e^{i\Lambda/2}\left[
                  \begin{array}{cc}
                    \cos(\Theta/2)e^{i(\Psi+\Phi)/2} & \sin(\Theta/2)e^{i(\Psi-\Phi)/2} \\
                    -\sin(\Theta/2)e^{i(\Psi-\Phi)/2} & \cos(\Theta/2)e^{-i(\Psi+\Phi)/2} \\
                  \end{array}
                \right]
\end{equation}
with the real parameters $\Theta$, $\Psi$, $\Phi$ and $\Lambda$ can be recovered by the above form.

By the concatenation and series product, the whole system $\mathcal{G}$ is given by
\begin{equation}
\mathcal{G}=G_3\vartriangleleft[(G_1\vartriangleleft M)\boxplus G_2]=(S_{\rm total},L_{\rm total},H_{\rm total}),
\end{equation}
where $S_{\rm total}=S_b\left[
                \begin{array}{cc}
                  s_{11}S & s_{12} \\
                  s_{21}S & s_{22} \\
                \end{array}
              \right]
$, $L_{\rm total}=\left[
          \begin{array}{c}
            s_{11}(L+SL_M) \\
            s_{21}(L+SL_M) \\
          \end{array}
        \right]
$, $H_{\rm total}=H+\lambda(t)\mathrm{Im}\{L^{\dag}S\sigma_-\}$.

Furthermore, the Lindblad superoperator $\mathcal{L}_{\mathcal{G}}(A\otimes X)$ for the whole system $\mathcal{G}$ can be expressed in the following form
\begin{equation}\label{supero}\begin{aligned}
\mathcal{L}_{\mathcal{G}}(A\otimes X)=&A\otimes\mathcal{L}_{G}X+(\mathcal{D}_{L_M}A)\otimes X+
L^\dag_MA\otimes S^\dag[X,L]\\
&+AL_M\otimes[L^\dag,X]S+L^\dag_MAL_M\otimes(S^{\dag}XS-X),
\end{aligned}\end{equation}
where $A$ is any operator of ancilla and $X$ is the system operator.

In what follows, we denote by $B_{i,t}$, which is a vacuum state, the input of signal model $M$ and $B_{v,t}$ is the vacuum noise for system $G_2$, then the total input, together with gauge process for the whole system $\mathcal{G}$ are given by
\begin{equation}\nonumber\begin{aligned}
B(t)=\left[
      \begin{array}{c}
        B_{i,t} \\
        B_{v,t} \\
      \end{array}
    \right], \Lambda(t)=\left[
                         \begin{array}{cc}
                           \Lambda_{i,t} & \Lambda_{iv,t} \\
                           \Lambda_{vi,t} & \Lambda_{v,t} \\
                         \end{array}
                       \right].
\end{aligned}\end{equation}
By the evolution
\begin{equation}\label{outputevol}\begin{aligned}
dB_{\mathrm{out}}(t)=&S_{\rm total}(t)dB(t)+L_{\rm total}(t)dt,\\
d\Lambda_{\mathrm{out}}(t)=&S_{\rm total}^\ast(t)d\Lambda(t)S_{\rm total}^T(t)+S_{\rm total}^\ast(t)dB^\ast(t)L_{\rm total}^T(t)
+L_{\rm total}^\ast(t)dB^T(t)S_{\rm total}^T(t)+L_{\rm total}^\ast(t)L_{\rm total}^T(t)dt,
\end{aligned}\end{equation}
we can get the output filed of the system
\begin{equation}\label{output}
dB_{\mathrm{out}}=\left[
                \begin{array}{c}
                  s_{11}SdB_{i,t}+s_{12}dB_{v,t}+s_{11}(L+SL_M)dt \\
                  s_{21}SdB_{i,t}+s_{22}dB_{v,t}+s_{21}(L+SL_M)dt \\
                \end{array}
              \right].
\end{equation}

\subsubsection{Q-P quadrature form}

Firstly, let $F=\left[
                  \begin{array}{cc}
                    1 & 0 \\
                    0 & -i \\
                  \end{array}
                \right]
$ and $G=0$, we use the quadrature form of the measurements, i.e.,
\begin{equation}\nonumber
dY_{1,t}=dB_{1,\mathrm{out}}+dB_{1,\mathrm{out}}^\ast,~~dY_{2,t}=\frac{dB_{2,\mathrm{out}}-dB_{2,\mathrm{out}}^\ast}{i}.
\end{equation}
Inserting the explicit form of system output \eqref{output}, we can get the measurements stochastic equations
\begin{eqnarray}\label{measurequ}\begin{aligned}
dY_{1,t}=&[s_{11}^\ast(L^\dag+L_M^\dag S^\dag)+s_{11}(L+SL_M)]dt+s_{11}^\ast S^\dag dB_{i,t}^\dag+s_{11}SdB_{i,t}+s_{12}^\ast dB_{v,t}^\dag+s_{12}dB_{v,t},\\
dY_{2,t}=&[is_{21}^\ast(L^\dag+L_M^\dag S^\dag)-is_{21}(L+SL_M)]dt+is_{21}^\ast S^\dag dB_{i,t}^\dag-is_{21}SdB_{i,t}+is_{22}^\ast dB_{v,t}^\dag-is_{22}dB_{v,t},
\end{aligned}\end{eqnarray}
where $dY_{1,t}$ and $dY_{2,t}$ are the homodyne detection measurements in the first and second channels, respectively. It can be verified that the expectation of the measurements are satisfied
\begin{equation}\begin{aligned}
&\tilde{\pi}_t(dY_{1,t})=\left[s^\ast_{11}\tilde{\pi}_t(L^\dag+L^\dag_MS^\dag)+s_{11}\tilde{\pi}_t(L+SL_M)\right]dt,\\
&\tilde{\pi}_t(dY_{2,t})=\left[is^\ast_{21}\tilde{\pi}_t(L^\dag+L^\dag_MS^\dag)-is_{21}\tilde{\pi}_t(L+SL_M)\right]dt,\\
&\tilde{\pi}_t(dY_{1,t}dY_{1,t})=\tilde{\pi}_t(dY_{2,t}dY_{2,t})=dt,\\
&\tilde{\pi}_t(dY_{1,t}dY_{2,t})=\tilde{\pi}_t(dY_{2,t}dY_{1,t})=0.
\end{aligned}\end{equation}

Then, the corresponding gain $\beta$ in Lemma \ref{peterlemma} can be given by
\begin{equation}\label{beta11}\begin{aligned}
\beta_1=&s_{11}\tilde{\pi}_t(A\otimes XL+AL_M\otimes XS)+s^\ast_{11}\tilde{\pi}_t(A\otimes L^\dag X+L^\dag_MA\otimes S^\dag X)\\
&-\tilde{\pi}_t(A\otimes X)[s^\ast_{11}\tilde{\pi}_t(L^\dag+L^\dag_MS^\dag)+s_{11}\tilde{\pi}_t(L+SL_M)],
\end{aligned}\end{equation}
\begin{equation}\label{beta12}\begin{aligned}
\beta_2=&-is_{21}\tilde{\pi}_t(A\otimes XL+AL_M\otimes XS)+is^\ast_{21}\tilde{\pi}_t(A\otimes L^\dag X+L^\dag_MA\otimes S^\dag X)\\
&-\tilde{\pi}_t(A\otimes X)[is^\ast_{21}\tilde{\pi}_t(L^\dag+L^\dag_MS^\dag)-is_{21}\tilde{\pi}_t(L+SL_M)],
\end{aligned}\end{equation}
where $X$ is the system operator and $A$ is any operator of the ancilla.

In what follows, we define \cite{GOUGH12QUANTUM}
\begin{equation}\begin{aligned}
\pi^{mn}_t(X)=\frac{\tilde{\pi}_t(A_{mn}\otimes X)}{d_{mn}},~~m,n=0,1,
\end{aligned}\end{equation}
where $A_{mn}$ and $d_{mn}$ have the following form
\begin{equation}\nonumber\begin{aligned}
A_{mn}=\left[
         \begin{array}{cc}
           A_{00} & A_{01} \\
           A_{10} & A_{11} \\
         \end{array}
       \right]=\left[
                 \begin{array}{cc}
                   \sigma_+\sigma_- & \sigma_+ \\
                   \sigma_- & I \\
                 \end{array}
               \right],~~
d_{mn}=\left[
         \begin{array}{cc}
           d_{00} & d_{01} \\
           d_{10} & d_{11} \\
         \end{array}
       \right]=\left[
                 \begin{array}{cc}
                   w(t) & \sqrt{w(t)} \\
                   \sqrt{w(t)} & 1 \\
                 \end{array}
               \right].
\end{aligned}\end{equation}
The following theorem gives the quantum filter for any quantum system driven by single-photon input field.
\begin{theorem}\label{theorem1}
Let $\{Y_{i,t},i=1,2\}$ be two homodyne detection measurements for a two-level quantum system driven by single-photon input field $|1_\xi\rangle$, the quantum filter for the conditional expectation is given by
\begin{scriptsize}
\begin{eqnarray}\label{qphei}\begin{aligned}
d\pi_t^{11}(X)=&\left\{\pi_t^{11}(\mathcal{L}_GX)+\pi_t^{01}(S^\dag[X,L])\xi^\ast(t)+\pi_t^{10}([L^\dag,X]S)\xi(t)
+\pi_t^{00}(S^\dag XS-X)|\xi(t)|^2\right\}dt\\
&+\Big[s_{11}\pi_t^{11}(XL)+s_{11}^\ast\pi_t^{11}(L^\dag X)+s_{11}\pi_t^{10}(XS)\xi(t)
+s_{11}^\ast\pi_t^{01}(S^\dag X)\xi^\ast(t)-\pi_t^{11}(X)k_1(t)\Big]dW_1(t)\\
&+\Big[-is_{21}\pi_t^{11}(XL)+is_{21}^\ast\pi_t^{11}(L^\dag X)-is_{21}\pi_t^{10}(XS)\xi(t)
+is_{21}^\ast\pi_t^{01}(S^\dag X)\xi^\ast(t)-\pi_t^{11}(X)k_2(t)\Big]dW_2(t),\\
d\pi_t^{10}(X)=&\left\{\pi_t^{10}(\mathcal{L}_GX)+\pi_t^{00}(S^\dag[X,L])\xi^\ast(t)\right\}dt\\
&+\left[s_{11}\pi_t^{10}(XL)+s_{11}^\ast\pi_t^{10}(L^\dag X)+s_{11}^\ast\pi_t^{00}(S^\dag X)\xi^\ast(t)-\pi_t^{10}(X)k_1(t)\right]dW_1(t)\\
&+\left[-is_{21}\pi_t^{10}(XL)+is_{21}^\ast\pi_t^{10}(L^\dag X)+is_{21}^\ast\pi_t^{00}(S^\dag X)\xi^\ast(t)-\pi_t^{10}(X)k_2(t)\right]dW_2(t),\\
d\pi_t^{00}(X)=&\pi_t^{00}(\mathcal{L}_GX)dt+\left[s_{11}\pi_t^{00}(XL)+s_{11}^\ast\pi_t^{00}(L^\dag X)-\pi_t^{00}(X)k_1(t)\right]dW_1(t)\\
&+\left[-is_{21}\pi_t^{00}(XL)+is_{21}^\ast\pi_t^{00}(L^\dag X)-\pi_t^{00}(X)k_2(t)\right]dW_2(t),
\end{aligned}\end{eqnarray}
\end{scriptsize}
where
\begin{eqnarray}\begin{aligned}
k_1(t)&=s_{11}^\ast\pi_t^{11}(L^\dag)+s_{11}\pi_t^{11}(L)+s_{11}^\ast\pi_t^{01}(S^\dag)\xi^\ast(t)+s_{11}\pi_t^{10}(S)\xi(t),\\
k_2(t)&=is_{21}^\ast\pi_t^{11}(L^\dag)-is_{21}\pi_t^{11}(L)+is_{21}^\ast\pi_t^{01}(S^\dag)\xi^\ast(t)-is_{21}\pi_t^{10}(S)\xi(t),
\end{aligned}\end{eqnarray}
and $\pi_t^{01}(X)=(\pi_t^{10}(X^\dagger))^\dagger$, the Wiener processes $W_1(t)$ and $W_2(t)$ are given by
\begin{equation}
dW_1(t)=dY_{1,t}-k_1(t)dt,~~dW_2(t)=dY_{2,t}-k_2(t)dt,
\end{equation}
respectively. The initial conditions are $\pi_0^{11}(X)=\pi_0^{00}(X)=\langle\eta,X\eta\rangle$, $\pi_0^{10}(X)=\pi_0^{01}(X)=0$.
\end{theorem}

If we write $\pi_t^{jk}(X)=\mathrm{Tr}[(\rho^{jk}(t))^\dag X]$, by the quantum filter \eqref{qphei}, we can get the following stochastic differential equations for the evolution of $\rho^{jk}(t)$.
\begin{corollary}\label{corollary1}
For a two-level quantum system driven by single-photon input field, the quantum filter in the Schr\"{o}dinger picture with two homodyne detection measurements is given by
\begin{scriptsize}
\begin{eqnarray}\label{qpsch}\begin{aligned}
d\rho^{11}(t)=&\left\{\mathcal{L}_G^\star\rho^{11}(t)+[S\rho^{01}(t),L^\dag]\xi(t)+[L,\rho^{10}(t)S^\dag]\xi^\ast(t)
+(S\rho^{00}(t)S^\dag-\rho^{00}(t))|\xi(t)|^2\right\}dt\\
&+\left[s_{11}^\ast\rho^{11}(t)L^\dag+s_{11}L\rho^{11}(t)+s_{11}^\ast\rho^{10}(t)S^\dag\xi^\ast(t)+
s_{11}S\rho^{01}(t)\xi(t)-\rho^{11}(t)k_1(t)\right]dW_1(t)\\
&+\left[is_{21}^\ast\rho^{11}(t)L^\dag-is_{21}L\rho^{11}(t)+is_{21}^\ast\rho^{10}(t)S^\dag\xi^\ast(t)-
is_{21}S\rho^{01}(t)\xi(t)-\rho^{11}(t)k_2(t)\right]dW_2(t),\\
d\rho^{10}(t)=&\left\{\mathcal{L}_G^\star\rho^{10}(t)+[S\rho^{00}(t),L^\dag]\xi(t)\right\}dt\\
&+\left[s_{11}^\ast\rho^{10}(t)L^\dag+s_{11}L\rho^{10}(t)+
s_{11}S\rho^{00}(t)\xi(t)-\rho^{10}(t)k_1(t)\right]dW_1(t)\\
&+\left[is_{21}^\ast\rho^{10}(t)L^\dag-is_{21}L\rho^{10}(t)-
is_{21}S\rho^{00}(t)\xi(t)-\rho^{10}(t)k_2(t)\right]dW_2(t),\\
d\rho^{00}(t)=&\mathcal{L}_G^\star\rho^{00}(t)dt+\left[s_{11}^\ast\rho^{00}(t)L^\dag+s_{11}L\rho^{00}(t)-\rho^{00}(t)k_1(t)\right]dW_1(t)\\
&+\left[is_{21}^\ast\rho^{00}(t)L^\dag-is_{21}L\rho^{00}(t)-\rho^{00}(t)k_2(t)\right]dW_2(t),
\end{aligned}\end{eqnarray}
\end{scriptsize}
where
\begin{eqnarray}\begin{aligned}
k_1(t)=&s_{11}\mathrm{Tr}[L\rho^{11}(t)]+s_{11}^\ast\mathrm{Tr}[L^\dag\rho^{11}(t)]+s_{11}\mathrm{Tr}[S\rho^{01}(t)]\xi(t)+s_{11}^\ast\mathrm{Tr}[S^\dag\rho^{10}(t)]\xi^\ast(t),\\
k_2(t)=&-is_{21}\mathrm{Tr}[L\rho^{11}(t)]+is_{21}^\ast\mathrm{Tr}[L^\dag\rho^{11}(t)]-is_{21}\mathrm{Tr}[S\rho^{01}(t)]\xi(t)+is_{21}^\ast\mathrm{Tr}[S^\dag\rho^{10}(t)]\xi^\ast(t),
\end{aligned}\end{eqnarray}
and $\rho^{01}(t)=(\rho^{10}(t))^\dagger$, the initial conditions are $\rho^{11}(0)=\rho^{00}(0)=|\eta\rangle\langle\eta|$, $\rho^{10}(0)=\rho^{01}(0)=0$.
\end{corollary}

\begin{remark}
If we let the beam splitter be
\begin{equation}\label{sb2}
S_b=\left[
      \begin{array}{cc}
        \sqrt{1-r^2}e^{i\theta} & re^{i\theta} \\
        -re^{i\theta} & \sqrt{1-r^2}e^{i\theta} \\
      \end{array}
    \right],~~0\leq r\leq1,
\end{equation}
the filtering equations \eqref{qphei} and \eqref{qpsch} will be reduced to the forms $(3.12)$ and $(3.13)$ in \cite{dong2015quantum}.
\end{remark}

\subsubsection{Q-Q quadrature form}

In this case, let $F=I$, $G=0$, we consider the two homodyne detection measurements have the following form
\begin{equation}\nonumber\begin{aligned}
dY_{1,t}=dB_{1,\mathrm{out}}+dB^\ast_{1,\mathrm{out}},~~dY_{2,t}=dB_{2,\mathrm{out}}+dB^\ast_{2,\mathrm{out}}.
\end{aligned}\end{equation}
It can be checked that the measurements stochastic equations $dY_{1,t}$ has the same form in \eqref{measurequ}, while
\begin{equation}\nonumber\begin{aligned}
dY_{2,t}=[s_{21}^\ast(L^\dag+L_M^\dag S^\dag)+s_{21}(L+SL_M)]dt+s_{21}^\ast S^\dag dB_{i,t}^\dag+s_{21}SdB_{i,t}+s_{22}^\ast dB_{v,t}^\dag+s_{22}dB_{v,t},
\end{aligned}\end{equation}
and the expectation of the measurements are also satisfied
\begin{equation}\nonumber\begin{aligned}
\tilde{\pi}_t(dY_tdY^T_t)=\left[
                            \begin{array}{cc}
                              dt & 0 \\
                              0 & dt \\
                            \end{array}
                          \right].
\end{aligned}\end{equation}
Furthermore, the corresponding gain $\beta$ in this case are given by
\begin{equation}\label{beta22}\begin{aligned}
\beta_2=&s_{21}\tilde{\pi}_t(A\otimes XL+AL_M\otimes XS)+s^\ast_{21}\tilde{\pi}_t(A\otimes L^\dag X+L^\dag_MA\otimes S^\dag X)\\
&-\tilde{\pi}_t(A\otimes X)[s^\ast_{21}\tilde{\pi}_t(L^\dag+L^\dag_MS^\dag)+s_{21}\tilde{\pi}_t(L+SL_M)],
\end{aligned}\end{equation}
and $\beta_1$ is as same as \eqref{beta11}.

Similarly, we can obtain the theorem which presents the quantum filter with Q-Q homodyne detection measurements. An alternative system of differential equations for the quantum filter in this case can be directly presented by letting the beam splitter parameter $s_{21}$ in \eqref{qphei} and \eqref{qpsch} be $is_{21}$. The explicit forms of quantum filtering equations in the Heisenberg picture and Schr\"{o}dinger picture can be seen in section \ref{appendix} for details.

\subsection{Quantum filter for homodyne detection plus photon-counting measurements}\label{homodyne+photon-counting}

In this subsection, we choose
\begin{equation}\nonumber\begin{aligned}
F=\left[
    \begin{array}{cc}
      1 & 0 \\
      0 & 0 \\
    \end{array}
  \right],~~G=\left[
    \begin{array}{cc}
      0 & 0 \\
      0 & 1 \\
    \end{array}
  \right],
\end{aligned}\end{equation}
then the filtering equations for a combination of homodyne detection and photon-counting measurements is presented. We can derive the measurements stochastic equations $dY_{1,t}$, which is as same as \eqref{measurequ}, and
\begin{equation}\label{4_1dy2}\begin{aligned}
dY_{2,t}=&s^\ast_{21}s_{21}(L^\dag+L^\dag_MS^\dag)(L+SL_M)dt\\
&+s^\ast_{21}s_{21}S(L^\dag+L^\dag_MS^\dag)dB_{i,t}+s^\ast_{21}s_{21}(L+SL_M)S^\dag dB^\dag_{i,t}\\
&+s^\ast_{21}s_{22}(L^\dag+L^\dag_MS^\dag)dB_{v,t}
+s^\ast_{22}s_{21}(L+SL_M)dB^\dag_{v,t}+dt.
\end{aligned}\end{equation}
In what follows, we assume that $S=I$, then the expectation of the measurements are satisfied
\begin{equation}\nonumber\begin{aligned}
\tilde{\pi}_t(dY_tdY^T_t)=\left[
                            \begin{array}{cc}
                              dt & 0 \\
                              0 & s^\ast_{21}s_{21}\tilde{\pi}_t[(L^\dag+L^\dag_M)(L+L_M)]dt \\
                            \end{array}
                          \right],
\end{aligned}\end{equation}
and the corresponding gains $\beta_1$ is given by \eqref{beta11}, while
\begin{equation}\label{beta21}\begin{aligned}
\beta_2=\frac{\tilde{\pi}_t(A\otimes L^\dag XL+L^\dag_MA\otimes XL+AL_M\otimes L^{\dag}X
+L^\dag_MAL_M\otimes X)}{\tilde{\pi}_t[(L^\dag+L^\dag_M)(L+L_M)]}-\tilde{\pi}_t(A\otimes X).
\end{aligned}\end{equation}
The following theorem presents the filtering equations for quantum system driven by single-photon state with a combination of homodyne detection and photon-counting measurements.
\begin{theorem}\label{theorem3}
Let $\{Y_{i,t},i=1,2\}$ be a combination of homodyne detection and photon-counting measurements. With single-photon input field, the quantum filter for the conditional expectation in the Heisenberg picture is given by
\begin{scriptsize}
\begin{eqnarray}\label{hphei}\begin{aligned}
d\pi^{11}_t(X)=&\left\{\pi^{11}_t(\mathcal{L}_GX)+\pi^{01}_t([X,L])\xi^\ast(t)+\pi^{10}_t([L^\dag,X])\xi(t)\right\}dt\\
&+\Big[s_{11}\pi^{11}_t(XL)+s_{11}^\ast\pi^{11}_t(L^{\dag}X)+s_{11}\pi^{10}_t(X)\xi(t)
+s_{11}^\ast\pi^{01}_t(X)\xi^\ast(t)-\pi^{11}_t(X)K_h(t)\Big]dW(t)\\
&+\Big\{K_p(t)^{-1}\left[\pi^{11}_t(L^\dag XL)+\pi^{01}_t(XL)\xi^\ast(t)+\pi^{10}_t(L^\dag X)\xi(t)+\pi^{00}_t(X)|\xi(t)|^2\right]
-\pi^{11}_t(X)\Big\}dN(t),\\
d\pi^{10}_t(X)=&\left\{\pi^{10}_t(\mathcal{L}_GX)+\pi^{00}_t([X,L])\xi^\ast(t)\right\}dt\\
&+\left[s_{11}\pi^{10}_t(XL)+s_{11}^\ast\pi^{10}_t(L^{\dag}X)+s_{11}^\ast\pi^{00}_t(X)\xi^\ast(t)-\pi^{10}_t(X)K_h(t)\right]dW(t)\\
&+\left\{K_p(t)^{-1}\left[\pi^{10}_t(L^\dag XL)+\pi^{00}_t(XL)\xi^\ast(t)\right]-\pi^{10}_t(X)\right\}dN(t),\\
d\pi^{00}_t(X)=&\pi^{00}_t(\mathcal{L}_GX)dt+\left[s_{11}\pi^{00}_t(XL)+s_{11}^\ast\pi^{00}_t(L^\dag X)-\pi^{00}_t(X)K_h(t)\right]dW(t)\\
&+\left\{K_p(t)^{-1}\left[\pi^{00}_t(L^\dag XL)\right]-\pi^{00}_t(X)\right\}dN(t),
\end{aligned}\end{eqnarray}
\end{scriptsize}
where
\begin{eqnarray}\begin{aligned}
K_h(t)&=s_{11}^\ast\pi_t^{11}(L^\dag)+s_{11}\pi_t^{11}(L)+s_{11}^\ast\pi_t^{01}(I)\xi^\ast(t)+s_{11}\pi_t^{10}(I)\xi(t),\\
K_p(t)&=\pi_t^{11}(L^\dag L)+\pi_t^{01}(L)\xi^\ast(t)+\pi_t^{10}(L^\dag)\xi(t)+\pi_t^{00}(I)|\xi(t)|^2,
\end{aligned}\end{eqnarray}
and $\pi^{01}_t(X)=(\pi^{10}_t(X^\dagger))^\dagger$, the Wiener process $W(t)$ and compensated Poisson process $N(t)$ are given by
\begin{equation}
dW(t)=dY_{1,t}-K_h(t)dt,~~dN(t)=dY_{2,t}-s^\ast_{21}s_{21}K_p(t)dt,
\end{equation}
respectively. The initial conditions are $\pi_0^{11}(X)=\pi_0^{00}(X)=\langle\eta,X\eta\rangle$, $\pi_0^{10}(X)=\pi_0^{01}(X)=0$.
\end{theorem}

Since we have got the filtering equations in the Heisenberg picture, by $\pi_t^{jk}(X)=\mathrm{Tr}[(\rho^{jk}(t))^\dag X]$, the evolution of the reduced density operator $\rho^{jk}(t)$ can be presented as follows.

\begin{corollary}\label{corollary3}
With a combination of homodyne detection and photon-counting measurements, the quantum filter for a two-level quantum system driven by single-photon input field in the Schr\"{o}dinger picture is given by
\begin{scriptsize}
\begin{eqnarray}\label{hpsch}\begin{aligned}
d\rho^{11}(t)=&\left\{\mathcal{L}^\star_G\rho^{11}(t)+[\rho^{01}(t),L^\dag]\xi(t)+[L,\rho^{10}(t)]\xi^\ast(t)\right\}dt\\
&+\left[s_{11}^\ast\rho^{11}(t)L^\dag+s_{11}L\rho^{11}(t)+s_{11}^\ast\rho^{10}(t)\xi^\ast(t)+s_{11}\rho^{01}(t)\xi(t)-K_h(t)\rho^{11}(t)\right]dW(t)\\
&+\left\{K_p(t)^{-1}\left[L\rho^{11}(t)L^\dag+\rho^{01}(t)L^\dag\xi(t)+L\rho^{10}(t)\xi^\ast(t)+\rho^{00}(t)|\xi(t)|^2\right]-\rho^{11}(t)\right\}dN(t),\\
d\rho^{10}(t)=&\left\{\mathcal{L}^\star_G\rho^{10}(t)+[\rho^{00}(t),L^\dag]\xi(t)\right\}dt\\
&+\left[s_{11}^\ast\rho^{10}(t)L^\dag+s_{11}L\rho^{10}(t)+s_{11}\rho^{00}(t)\xi(t)-K_h(t)\rho^{10}(t)\right]dW(t)\\
&+\left\{K_p(t)^{-1}\left[L\rho^{10}(t)L^\dag+\rho^{00}(t)L^\dag\xi(t)\right]-\rho^{10}(t)\right\}dN(t),\\
d\rho^{00}(t)=&\mathcal{L}^\star_G\rho^{00}(t)dt+\left[s_{11}^\ast\rho^{00}(t)L^\dag+s_{11}L\rho^{00}(t)-K_h(t)\rho^{00}(t)\right]dW(t)\\
&+\left\{K_p(t)^{-1}\left[L\rho^{00}(t)L^\dag\right]-\rho^{00}(t)\right\}dN(t),
\end{aligned}\end{eqnarray}
\end{scriptsize}
where
\begin{eqnarray}\begin{aligned}
K_h(t)=&s_{11}\mathrm{Tr}[L\rho^{11}(t)]+s_{11}^\ast\mathrm{Tr}[L^\dag\rho^{11}(t)]+s_{11}\mathrm{Tr}[\rho^{01}(t)]\xi(t)+s_{11}^\ast\mathrm{Tr}[\rho^{10}(t)]\xi^\ast(t),\\
K_p(t)=&\mathrm{Tr}[L^\dag L\rho^{11}(t)]+\mathrm{Tr}[L^\dag\rho^{01}(t)]\xi(t)+\mathrm{Tr}[L\rho^{10}(t)]\xi^\ast(t)+\mathrm{Tr}[\rho^{00}(t)]|\xi(t)|^2,
\end{aligned}\end{eqnarray}
and $\rho^{01}(t)=(\rho^{10}(t))^\dagger$, the initial conditions are $\rho^{11}(0)=\rho^{00}(0)=|\eta\rangle\langle\eta|$, $\rho^{10}(0)=\rho^{01}(0)=0$.
\end{corollary}

\begin{remark}
Similarly, let
\begin{equation}\nonumber
S_b=\left[
                               \begin{array}{cc}
                                 \sqrt{1-r^2}e^{i\theta} & re^{i(\theta+\frac{\pi}{2})} \\
                                 re^{i(\theta+\frac{\pi}{2})} & \sqrt{1-r^2}e^{i\theta} \\
                               \end{array}
                             \right],~~0\leq r\leq1,
\end{equation}
the filtering equations \eqref{hphei} and \eqref{hpsch} will be equivalent to the forms $(3.8)$ and $(3.9)$ ($S=I$) which have been presented in \cite{dong2015quantum}.
\end{remark}

\subsection{Quantum filter for two photon-counting measurements}\label{twophoton-counting}

In this subsection, we choose
\begin{equation}\nonumber\begin{aligned}
F=0,~~G=I,
\end{aligned}\end{equation}
then the filtering equations for two photon-counting measurements is given. We can get the measurements stochastic equations $dY_{2,t}$, which is as same as \eqref{4_1dy2}, and
\begin{equation}\nonumber\begin{aligned}
dY_{1,t}=&s^\ast_{11}s_{11}(L^\dag+L^\dag_MS^\dag)(L+SL_M)dt\\
&+s^\ast_{11}s_{11}S(L^\dag+L^\dag_MS^\dag)dB_{i,t}+s^\ast_{11}s_{11}(L+SL_M)S^\dag dB^\dag_{i,t}\\
&+s^\ast_{11}s_{12}(L^\dag+L^\dag_MS^\dag)dB_{v,t}
+s^\ast_{12}s_{11}(L+SL_M)dB^\dag_{v,t}+dt.
\end{aligned}\end{equation}
Furthermore, we also assume $S=I$, then the expectation of the measurements are satisfied
\begin{equation}\nonumber\begin{aligned}
\tilde{\pi}_t(dY_tdY^T_t)=\left[
                            \begin{array}{cc}
                              s^\ast_{11}s_{11} & 0 \\
                              0 & s^\ast_{21}s_{21} \\
                            \end{array}
                          \right]\tilde{\pi}_t[(L^\dag+L^\dag_M)(L+L_M)]dt.
\end{aligned}\end{equation}
By simple calculation, we can find that $\beta_1=\beta_2$ in \eqref{beta21}. It means that the corresponding gain of photon-counting detection is independent with beam splitter parameters.
The following theorem presents the filtering equations for quantum system driven by single-photon state with two photon-counting detection measurements.

\begin{theorem}\label{theorem4}
Let $\{Y_{i,t},i=1,2\}$ be two photon-counting measurements. With single-photon input field, the quantum filter for the conditional expectation in the Heisenberg picture is given by
\begin{scriptsize}
\begin{eqnarray}\label{pphei}\begin{aligned}
d\pi^{11}_t(X)=&\left\{\pi^{11}_t(\mathcal{L}_GX)+\pi^{01}_t([X,L])\xi^\ast(t)+\pi^{10}_t([L^\dag,X])\xi(t)\right\}dt\\
&+\sum_{i=1}^2\bigg\{K_p(t)^{-1}\left[\pi^{11}_t(L^\dag XL)+\pi^{01}_t(XL)\xi^\ast(t)+
\pi^{10}_t(L^\dag X)\xi(t)+\pi^{00}_t(X)|\xi(t)|^2\right]
-\pi^{11}_t(X)\bigg\}dN_i(t),\\
d\pi^{10}_t(X)=&\left\{\pi^{10}_t(\mathcal{L}_GX)+\pi^{00}_t([X,L])\xi^\ast(t)\right\}dt\\
&+\sum_{i=1}^2\left\{K_p(t)^{-1}\left[\pi^{10}_t(L^\dag XL)+\pi^{00}_t(XL)\xi^\ast(t)\right]-\pi^{10}_t(X)\right\}dN_i(t),\\
d\pi^{00}_t(X)=&\pi^{00}_t(\mathcal{L}_GX)dt+\sum_{i=1}^2\left\{K_p(t)^{-1}\left[\pi^{00}_t(L^\dag XL)\right]-\pi^{00}_t(X)\right\}dN_i(t),
\end{aligned}\end{eqnarray}
\end{scriptsize}
where
\begin{eqnarray}\begin{aligned}
K_p(t)=\pi_t^{11}(L^\dag L)+\pi_t^{01}(L)\xi^\ast(t)+\pi_t^{10}(L^\dag)\xi(t)+\pi_t^{00}(I)|\xi(t)|^2,
\end{aligned}\end{eqnarray}
and $\pi^{01}_t(X)=(\pi^{10}_t(X^\dagger))^\dagger$, the compensated Poisson processes $N_i(t)$ are given by
\begin{equation}
dN_1(t)=dY_{1,t}-s^\ast_{11}s_{11}K_p(t)dt,~~dN_2(t)=dY_{2,t}-s^\ast_{21}s_{21}K_p(t)dt,
\end{equation}
respectively. The initial conditions are $\pi_0^{11}(X)=\pi_0^{00}(X)=\langle\eta,X\eta\rangle$, $\pi_0^{10}(X)=\pi_0^{01}(X)=0$.
\end{theorem}

By $\pi_t^{jk}(X)=\mathrm{Tr}[(\rho^{jk}(t))^\dag X]$, the evolution of the reduced density operator $\rho^{jk}(t)$ can be presented directly.

\begin{corollary}\label{corollary4}
With two photon-counting measurements, the quantum filter for a two-level quantum system driven by single-photon input field in the Schr\"{o}dinger picture is given by
\begin{scriptsize}
\begin{eqnarray}\label{ppsch}\begin{aligned}
d\rho^{11}(t)=&\left\{\mathcal{L}^\star_G\rho^{11}(t)+[\rho^{01}(t),L^\dag]\xi(t)+[L,\rho^{10}(t)]\xi^\ast(t)\right\}dt\\
&+\sum_{i=1}^2\bigg\{K_p(t)^{-1}\left[L\rho^{11}(t)L^\dag+\rho^{01}(t)L^\dag\xi(t)+L\rho^{10}(t)\xi^\ast(t)+\rho^{00}(t)|\xi(t)|^2\right]
-\rho^{11}(t)\bigg\}dN_i(t),\\
d\rho^{10}(t)=&\left\{\mathcal{L}^\star_G\rho^{10}(t)+[\rho^{00}(t),L^\dag]\xi(t)\right\}dt\\
&+\sum_{i=1}^2\left\{K_p(t)^{-1}\left[L\rho^{10}(t)L^\dag+\rho^{00}(t)L^\dag\xi(t)\right]-\rho^{10}(t)\right\}dN_i(t),\\
d\rho^{00}(t)=&\mathcal{L}^\star_G\rho^{00}(t)dt+\sum_{i=1}^2\left\{K_p(t)^{-1}\left[L\rho^{00}(t)L^\dag\right]-\rho^{00}(t)\right\}dN_i(t),
\end{aligned}\end{eqnarray}
\end{scriptsize}
where
\begin{eqnarray}\begin{aligned}
K_p(t)=\mathrm{Tr}[L^\dag L\rho^{11}(t)]+\mathrm{Tr}[L^\dag\rho^{01}(t)]\xi(t)+\mathrm{Tr}[L\rho^{10}(t)]\xi^\ast(t)+\mathrm{Tr}[\rho^{00}(t)]|\xi(t)|^2,
\end{aligned}\end{eqnarray}
and $\rho^{01}(t)=(\rho^{10}(t))^\dagger$, the initial conditions are $\rho^{11}(0)=\rho^{00}(0)=|\eta\rangle\langle\eta|$, $\rho^{10}(0)=\rho^{01}(0)=0$.
\end{corollary}

Quantum filters describe the joint system-field dynamics conditioned on measurement outputs, while master equations describe the system dynamics itself.  In this sense, master equations can be regarded as unconditional system dynamics, see e.g., \cite{breuer2002theory,barchielli2009quantum,wiseman2009quantum,GOUGH12QUANTUM}. In this paper, the master equations we used are Lindblad master equations (also called ensemble average dynamics), which can be directly obtained by tracing out the field from the filtering equations in any case above. In what follows, the master equations are explicitly presented in the Schr\"{o}dinger picture

\begin{eqnarray}\label{mesch}\begin{aligned}
\dot{\rho}^{11}(t)=&\mathcal{L}^\star_G\rho^{11}(t)+[\rho^{01}(t),L^\dag]\xi(t)+[L,\rho^{10}(t)]\xi^\ast(t),\\
\dot{\rho}^{10}(t)=&\mathcal{L}^\star_G\rho^{10}(t)+[\rho^{00}(t),L^\dag]\xi(t),\\
\dot{\rho}^{00}(t)=&\mathcal{L}^\star_G\rho^{00}(t),
\end{aligned}\end{eqnarray}

where $\rho^{01}(t)=(\rho^{10}(t))^\dagger$, the initial conditions are $\rho^{11}(0)=\rho^{00}(0)=|\eta\rangle\langle\eta|$, $\rho^{10}(0)=\rho^{01}(0)=0$.

\subsection{Simulation results}\label{simulation}

The problem of how to efficiently excite a two-level atom by a single photon has been investigated in, e.g., \cite{Gheri1998photon,loudon2000quantum,MGG09,RSF10,wang2011efficient,GOUGH12QUANTUM,GJN13}. For example, if the photon has an exponentially decaying pulse shape with the decay rate equal to the coupling strength, then the two-level atom can be fully excited \cite{MGG09, wang2011efficient}. On the other hand, if the photon has a Gaussian pulse shape, the maximum excitation rate is 0.8 \cite{wang2011efficient,GOUGH12QUANTUM,SONG13MULTI}. In our simulation, we apply the filtering equations presented in subsections \ref{twohomodyne} and \ref{homodyne+photon-counting} to the problem of exciting a two-level atom by a single photon in Gaussian pulse shape. This system can be parameterized as follows. The scattering operator and coupling operator are $S=I$, $L=\kappa\sigma_-$, respectively. The atom is supposed to be in the ground state $|g\rangle\langle g|$ initially with the Hamiltonian $H=0$. The input pulse shape $\xi(t)$ for the single photon is given by
\begin{equation}\label{pulse}\begin{aligned}
\xi(t)=\left(\frac{\Omega^2}{2\pi}\right)^{1/4}\exp\left[-\frac{\Omega^2}{4}(t-t_0)^2\right],
\end{aligned}\end{equation}
where $t_0$ is the peak arrival time of the wave packet and $\Omega$ is the frequency bandwidth.

In what follows, we choose the beam splitter $S_b=\left[
                             \begin{array}{cc}
                               \sqrt{1-r^2} & ir \\
                               ir & \sqrt{1-r^2} \\
                             \end{array}
                           \right]$,
the peak arrival time $t_0=3$, and the frequency bandwidth $\Omega=1.5\kappa$, which is the optimum pulse bandwidth to maximize the absorption \cite{wang2011efficient,GOUGH12QUANTUM}. We wish to calculate the excitation probability
\begin{equation}\nonumber\begin{aligned}
P_e(t)=\langle e|\rho^{11}(t)|e\rangle,
\end{aligned}\end{equation}
where $|e\rangle$ means the excited state, $\rho^{11}(t)$ is the solution to \eqref{qpsch}, \eqref{hpsch}, and \eqref{ppsch}, respectively.

\subsubsection{Two homodyne detection measurements}

\begin{figure}
\centering
\includegraphics[width=1.0\textwidth]{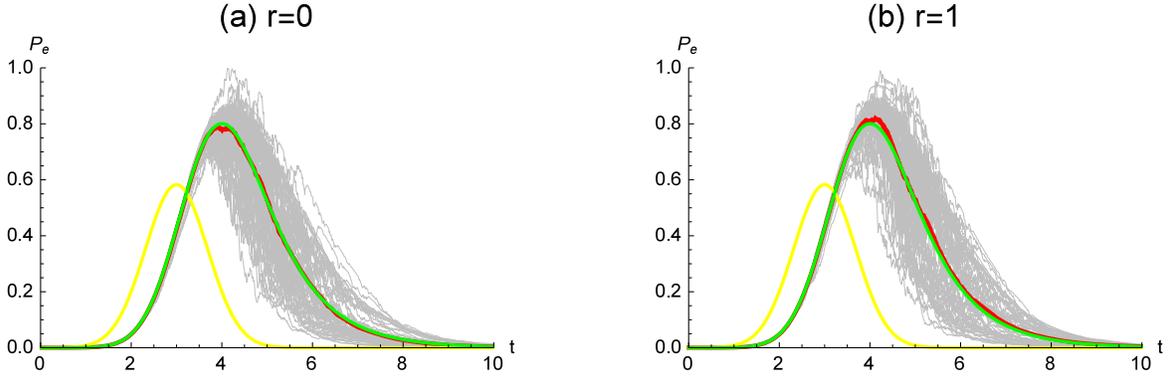}
\caption{\label{fig_5}(Color online) The excitation probability for a two-level atom driven by a single photon without any vacuum noise. The yellow curve is the wave packet $|\xi(t)|^2$, the green curve is $P_e(t)$ given by the master equation, the gray curves are the trajectories and the red curve denotes the average of these trajectories. Both the two cases for the beam splitter parameter (a) $r=0$ and (b) $r=1$ are equivalent to the ideal scenario in \cite{GOUGH12QUANTUM}.}
\end{figure}

Firstly, we consider the ideal case, which means that there is no any vacuum noise in the quantum system \cite{GOUGH12QUANTUM}. It can be achieved by letting the beam splitter parameter $r=0$ or $r=1$. In Fig. \ref{fig_5}, 64 different quantum trajectories are simulated as gray curves in each case. Particularly, Fig. \ref{fig_5} (a) ($r=0$) denotes the ideal case of no noise, which is equivalent to the first single homodyne detection, while Fig. \ref{fig_5} (b) ($r=1$) means that the single measurement turns out to be the second homodyne detection. We can see that many of the stochastic trajectories begin to decay after the main part of the wave packet, i.e., $t=4$. Meanwhile, approximately $33\%$ of the trajectories can rise beyond excitation probability $P_e(t)=0.9$, some of which continue to rise towards $P_e(t)=1$, it means that the atom may be fully excited.

\begin{figure}
\centering
\includegraphics[width=1.0\textwidth]{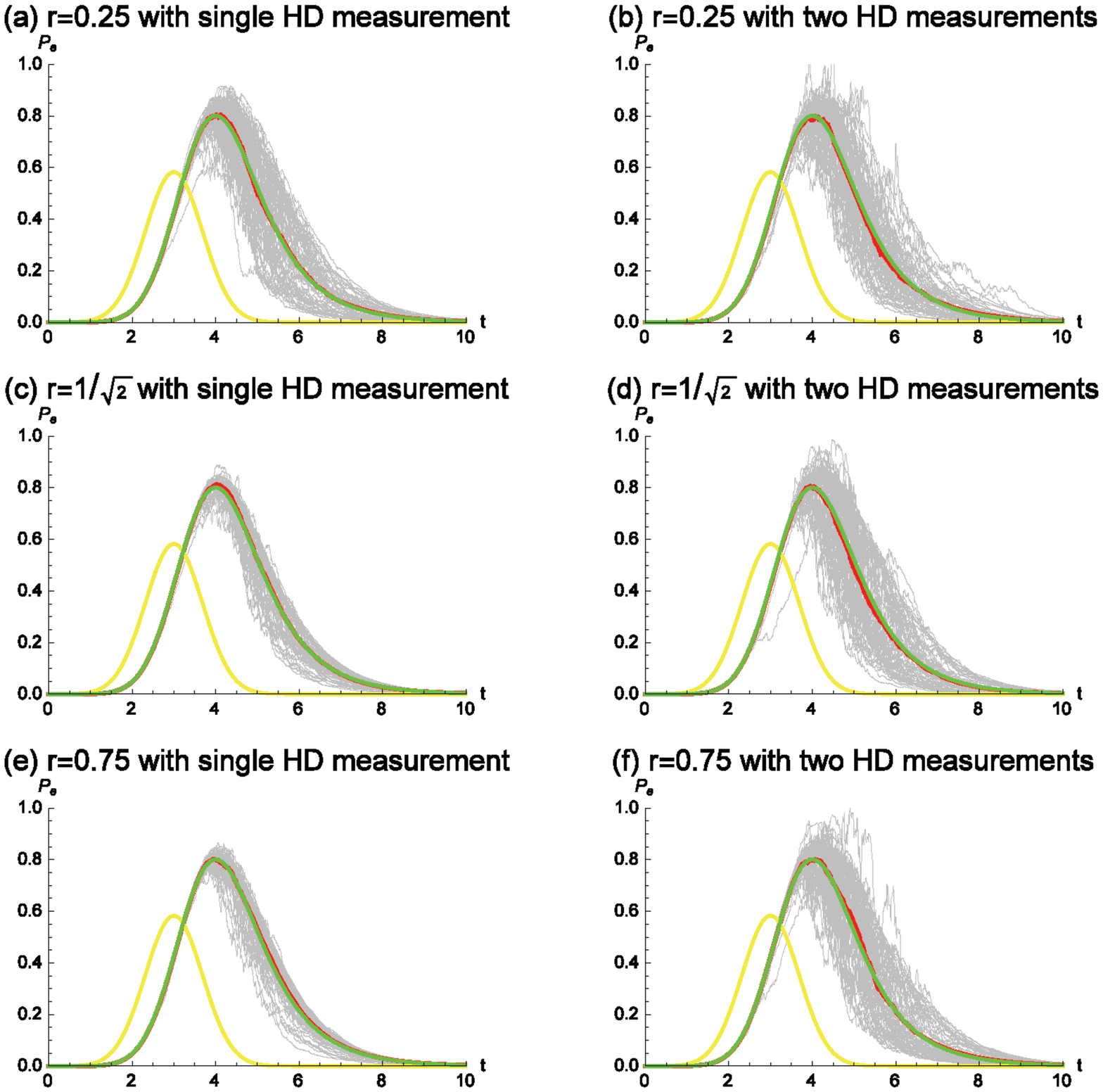}
\caption{\label{fig_6}(Color online) The excitation probability for a two-level atom driven by a single photon which is contaminated by quantum vacuum noise. Here we use (a)(c)(e) single homodyne detection measurement,
(b)(d)(f) two homodyne detection measurements, with different beam splitter parameters at the outputs of quantum system.}
\end{figure}

Then, the more realistic situation is considered in Fig. \ref{fig_6}. Since the output field state has been discussed before, i.e., $|\psi_{\mathrm{out}}\rangle=s_{11}|1_\eta\rangle\otimes|0\rangle+s_{21}|0\rangle\otimes|1_\eta\rangle$, we will see the drawback of general filtering method (single detection measurement) by comparing the simulation results. The two-level atom is contaminated by vacuum noise, and we use (a), (c), (e) single homodyne detection, (b), (d), (f) two homodyne detections at the outputs with different beam splitter parameters, respectively. That is, $P_e(t)$ in Fig. \ref{fig_6} (a), (c), (e) is the solution to the filtering equations \eqref{qpsch} with only one detection term $dW_1(t)$. While, in Fig. \ref{fig_6} (b), (d), (f) $P_e(t)$ is the solution with two detection terms $dW_1(t)$ and $dW_2(t)$. For the two homodyne detectors case, approximately $33\%$ of the trajectories can rise beyond the referred excitation probability $P_e(t)=0.9$, some of which can even rise up to $P_e(t)=1$. Thus, the performance of filtering setting with two homodyne detectors is equivalent to that in the ideal case in Fig. \ref{fig_5}. On the other hand, for the one homodyne detector, in our simulations there is no single trajectory which can go above $0.9$. Therefore, the advantage of multiple homodyne detection measurements is clear. In addition, measurement back-action can be reflected by comparing the simulation results between left and right sides in Fig. \ref{fig_6}. Although the systems and input fields are the same in the two cases, filtering results can be much different by selecting the measurements. This is also a distinct feature of quantum filtering from the Kalman filter in classical system. It is obvious that the efficiency of exciting an atom could be improved significantly by choosing multiple homodyne detection measurements.

\subsubsection{Homodyne detection plus photon-counting measurements}

\begin{figure}[!htb]
\centering
\begin{minipage}{5cm}
\centering
\includegraphics[width=1.0\textwidth]{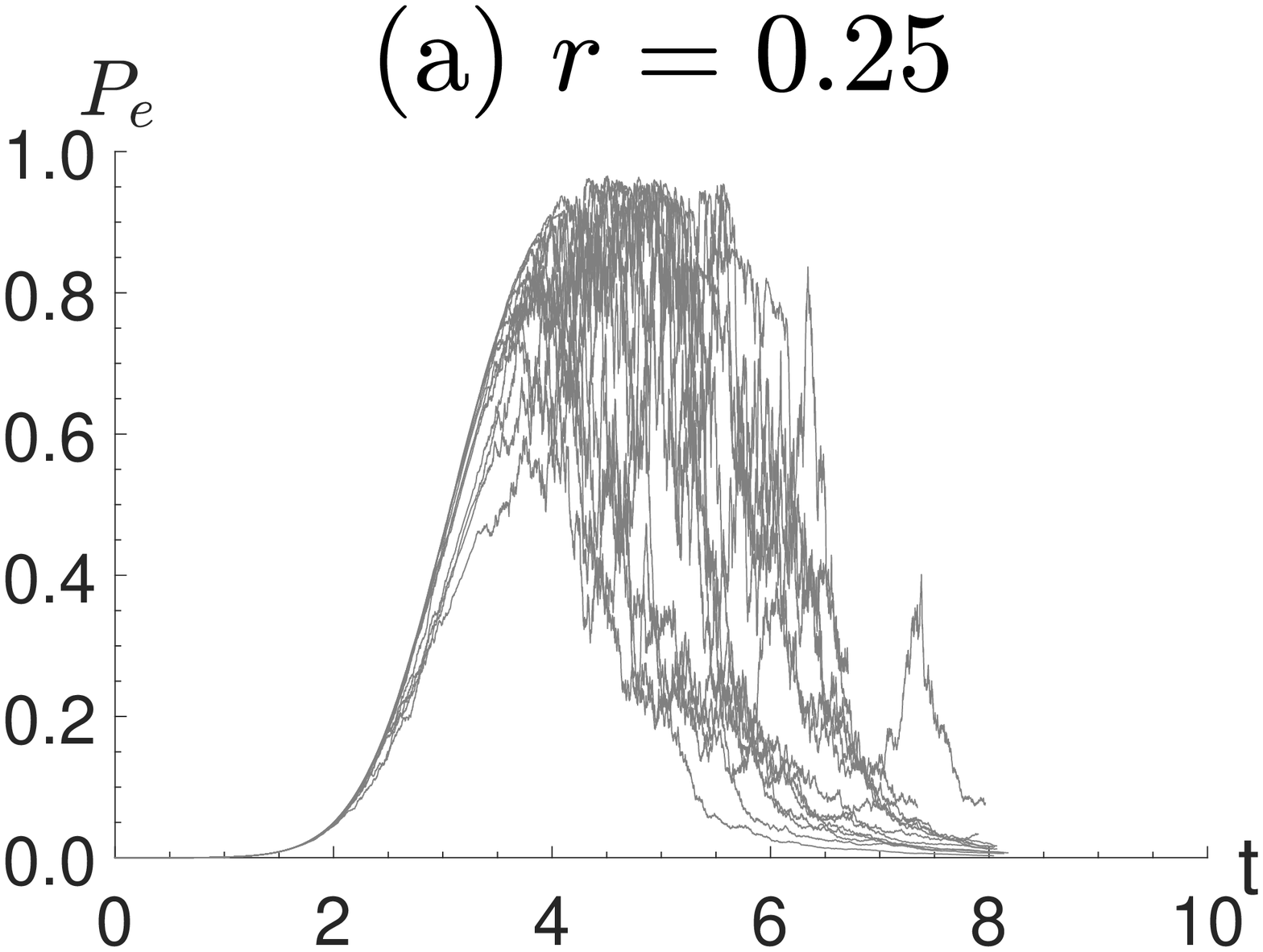}
\end{minipage}
\begin{minipage}{5cm}
\centering
\includegraphics[width=1.0\textwidth]{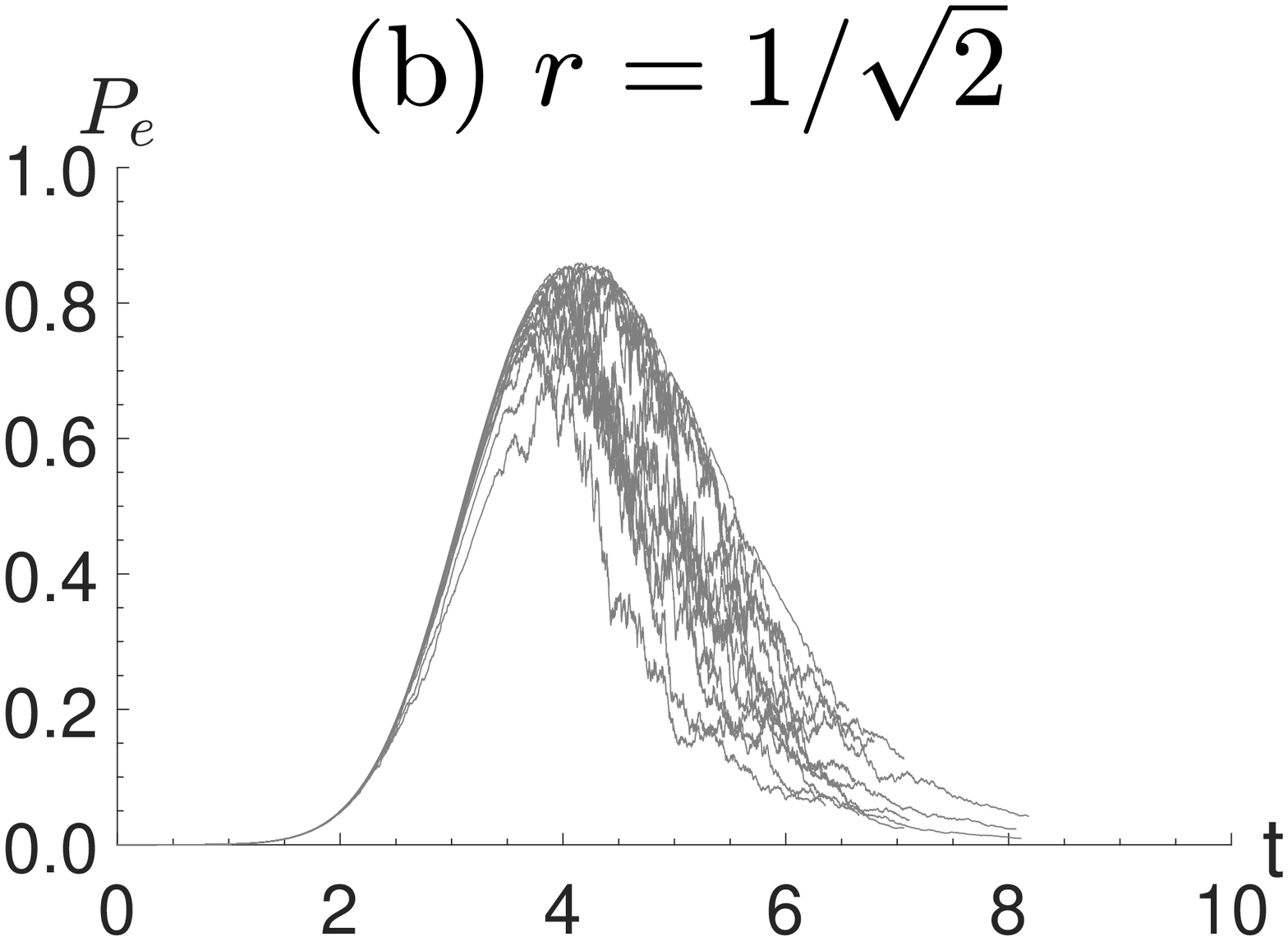}
\end{minipage}
\begin{minipage}{5cm}
\centering
\includegraphics[width=1.0\textwidth]{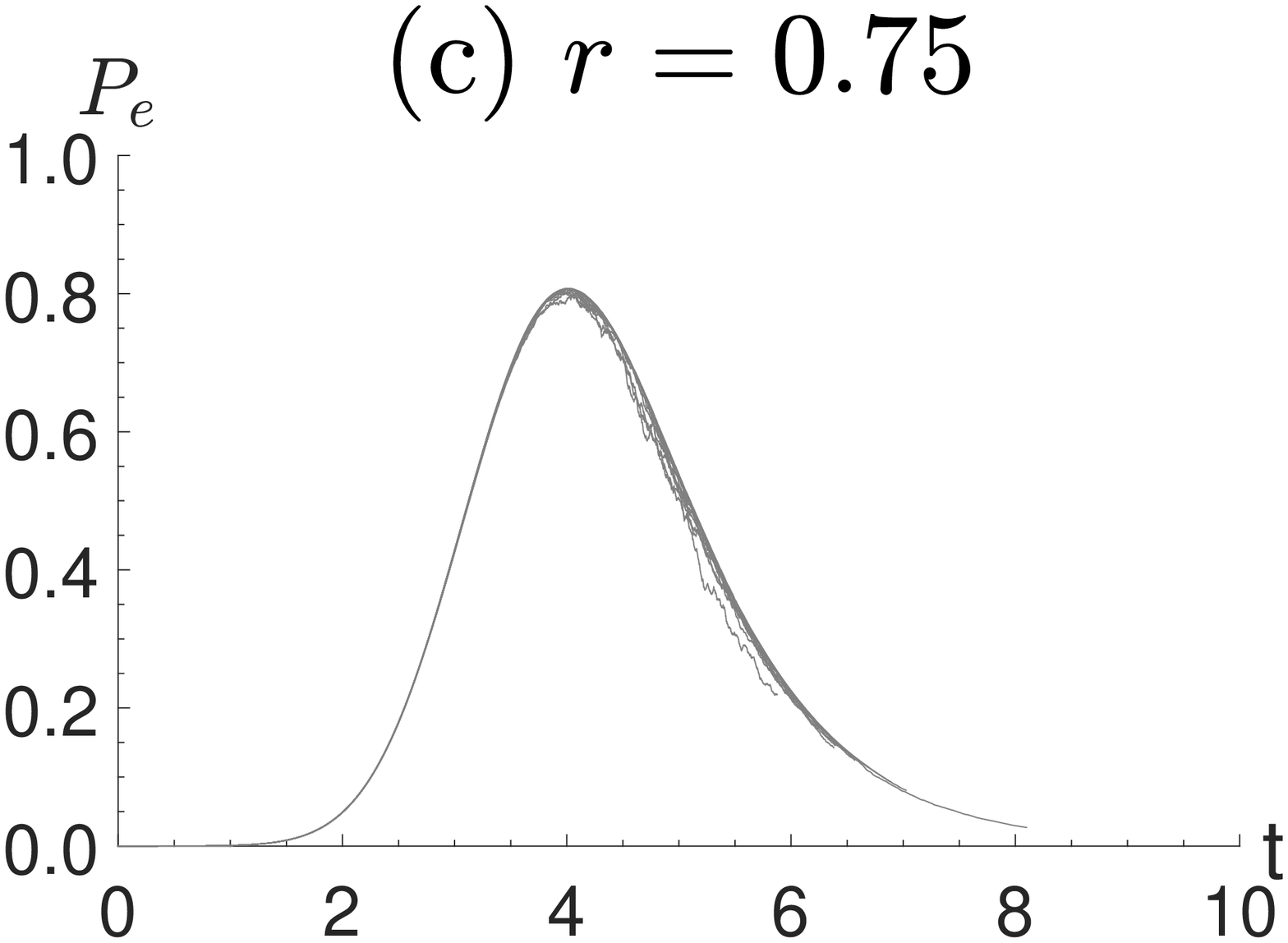}
\end{minipage}
\caption{\label{fig_7}(Color online) The excitation probability for a two-level atom driven by a single photon which is contaminated by quantum vacuum noise. Here we use homodyne detection plus photon-counting measurements, with different beam splitter parameters at the outputs of quantum system.}
\end{figure}

In this case, we consider that the detectors are the combined measurements of homodyne detection and photon-counting detection. That is, $P_e(t)$ in Fig. \ref{fig_7} (a), (b), and (c) is the solution to the filtering equations \eqref{hpsch} with homodyne detection term $dW(t)$ and photon-counting term $dN(t)$. Specially, we choose $r=0.25$ in Fig. \ref{fig_7} (a) which means that the homodyne detector is given more weight than the photo-detector. Compared with Fig. \ref{fig_6} (a), (c), and (e) (single homodyne detector), some trajectories can rise beyond $P_e(t)=0.9$, which reveals that multiple detection measurements is also better to excite a two-level atom. On the other hand, the wave-particle duality of light can be recovered by comparing Fig. \ref{fig_7} (a), (b), and (c). When the beam splitter parameter $r$ is increasing, the photo-detector (reflecting the particle nature) is given more and more weight than the homodyne detector (reflecting the wave nature). At the same time, fewer trajectories with high excitation probability can be obtained and all trajectories are concentrated at the master equation. To the best knowledge of the authors, this has never been reported in the literature.

\subsubsection{Two photon-counting measurements}

\begin{figure}[!htb]
\centering
\begin{minipage}{5cm}
\centering
\includegraphics[width=1.0\textwidth]{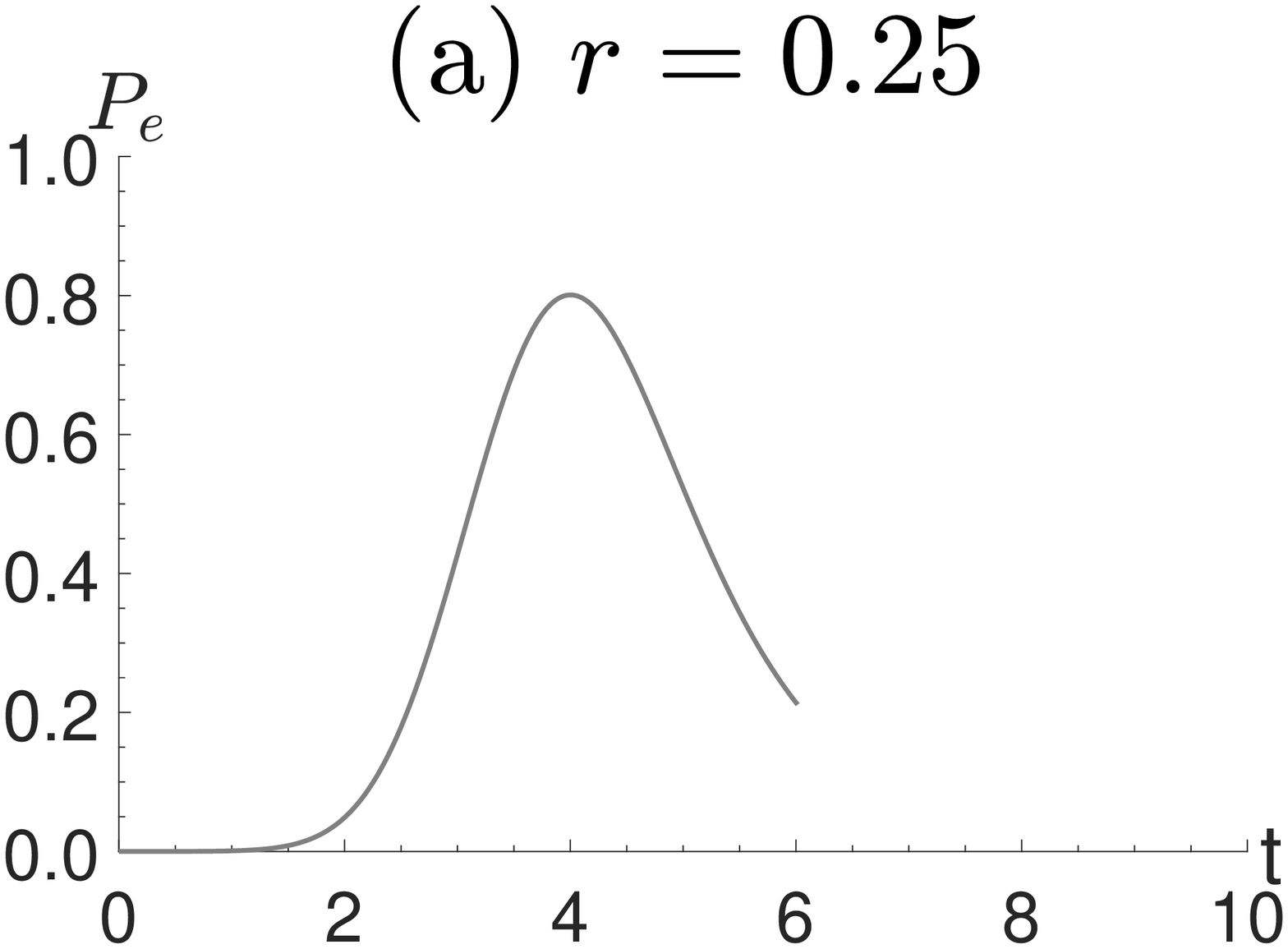}
\end{minipage}
\begin{minipage}{5cm}
\centering
\includegraphics[width=1.0\textwidth]{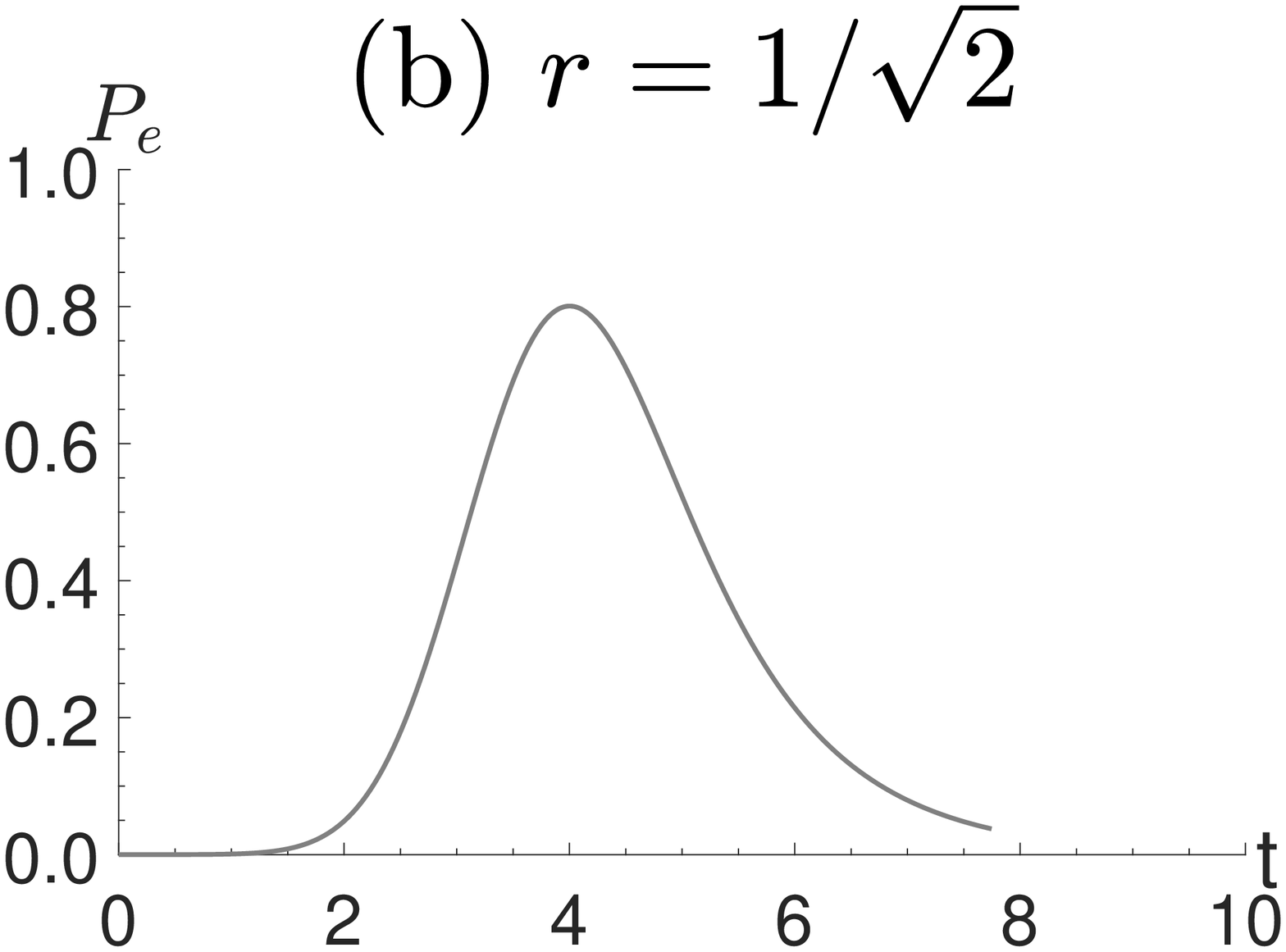}
\end{minipage}
\begin{minipage}{5cm}
\centering
\includegraphics[width=1.0\textwidth]{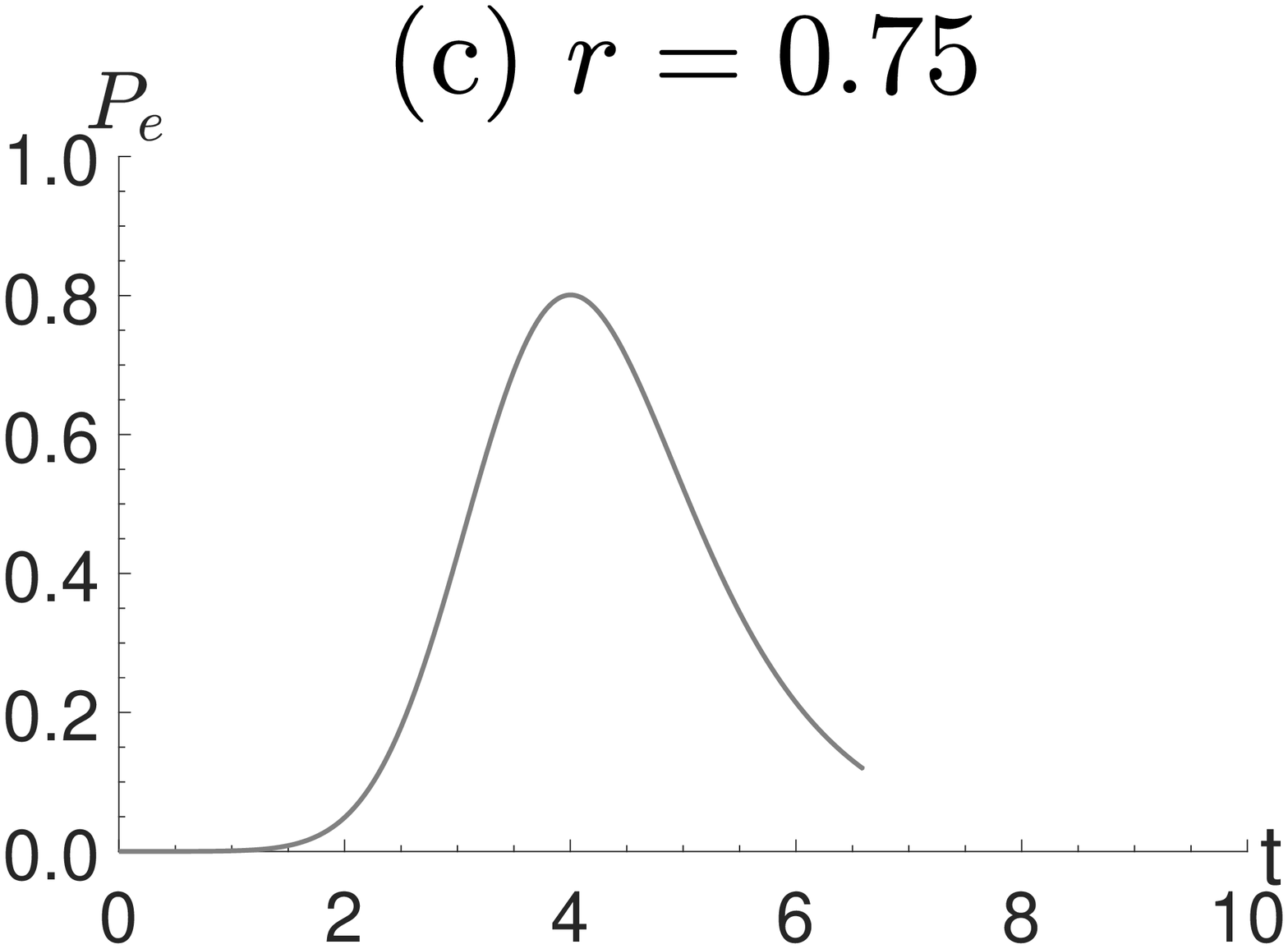}
\end{minipage}
\caption{\label{fig_8}(Color online) The excitation probability for a two-level atom driven by a single photon which is contaminated by quantum vacuum noise. Here we use two photon-counting measurements, with different beam splitter parameters at the outputs of quantum system.}
\end{figure}

In this case, we consider that the detectors are two photon-counting measurements. That is, $P_e(t)$ in Fig. \ref{fig_8} (a), (b), and (c) is the solution to the filtering equations \eqref{ppsch} with two photon-counting detection terms $dN_i(t)$, $i=1,2$. Specially, we also choose $r=0.25$, $r=1/\sqrt{2}$ and $r=0.75$ respectively in Fig. \ref{fig_8} (a), (b), and (c) to compare the specific quantum trajectories with different weights between the two measurement channels. In this filter setting, the excitation probability $P_e(t)$ is as same as the master equation before the photon is detected. This can be explained in the following way:  a master equation presents the ensemble average dynamics, which corresponds to an unconditional dynamics. For the two photon-counting case, before the detection of the photon, the process  is just an unconditional dynamics.   Thus, these two dynamics give the identical result  before the photon is detected. After the detection of the photon,  the excitation probability $P_e(t)$ vanishes  as the two-level atom is in the ground state. In this case, the filtering setting with multiple photon-counting measurements has no advantage over single photon-counting measurement. It also can be seen that the photon detection time $t$ is a random variable in Fig. \ref{fig_8}.

\begin{remark}
Based on the studies carried out in this paper, we can make the following conclusions.
\begin{itemize}
\item In the ideal case (namely, no additional quantum white noise), with single homodyne detection measurement in Fig. \ref{fig_5}, approximately $33\%$ of the trajectories can rise beyond excitation probability $P_e(t)=0.9$, some of which continue to rise towards $P_e(t)=1$, it means that the atom may be fully excited. This is consistent with the results in \cite{GOUGH12QUANTUM}.

\item In the ideal case (namely, no additional quantum white noise), with single photon-counting measurement, the excitation probability $P_e(t)$ can be at most $0.8$.

\item In the imperfect case (namely, under additional quantum white noise and mode mismatch):
\begin{enumerate}
\item If only one homodyne detector is used, there is no single trajectory which can go above $P_e(t)=0.9$, see Fig. \ref{fig_6} (a), (c), and (e).

\item If two homodyne detectors are used, approximately $33\%$ of the trajectories in Fig. \ref{fig_6}(b), (d), and (f) can rise beyond the referred excitation probability $P_e(t)=0.9$, some of which can even rise up to $P_e(t)=1$. That is, this filtering setting has the same performance as the ideal case in Fig. \ref{fig_5}.

\item In the homodyne detector plus photo-detector setting, when $r=0$, we get the ideal case with single homodyne detection measurement; when $r=1$, we get the ideal case with single photon-counting measurement. Specially, for sufficiently small $r$, the excitation performance is better than that in the single homodyne detection, compare Fig. \ref{fig_7}(a) with Fig. \ref{fig_6}(a), (c), (e). For bigger $r$, the excitation performance is worse than that in the single homodyne detection, compare Fig. \ref{fig_7}(c) with   Fig. \ref{fig_6}(a), (c), (e).

\item If two photo-detectors are used, the excitation probability $P_e(t)$ can be at most $0.8$. It is comparable to the ideal case with single photon-counting measurement. In this case, multiple measurements have no advantage over single measurement.
\end{enumerate}
\end{itemize}
\end{remark}

\section{Conclusions}\label{conclusions}

In this paper, the quantum filters for a two-level system driven by single-photon input state with multiple compatible measurements have been presented. Particularly, the explicit forms of filtering equations with two homodyne detection measurements (Q-P and Q-Q quadrature forms), a combination of homodyne detection and photon-counting, and two photon-counting detections are given. The numerical simulations of exciting a two-level atom driven by a single photon which is contaminated by quantum vacuum noise are conducted. Comparison of numerical results demonstrates the advantage of filter design with multiple measurements.

In the future work, we are considering the stability of single photon filtering. Our approach is based on applying the method applied in \cite{amini2014stability}. As a further direction, we can study the filtering problem when we consider the multi-photon input state \cite{SONG13MULTI}. Also, we may take into account imperfections in measurements. Moreover, showing the stability of multi-photon filtering is in the perspective of our research.

\section{Acknowledgements}
The authors would like to thank the anonymous reviewers for their detailed comments which helped to improve the quality of the paper. This work was financially supported in part by National Natural Science Foundation of China (NSFC) grant (No. 61374057), Hong Kong RGC grant (Nos. 531213 and 15206915), JCJC INS2I 2016 ``QIGR3CF'' project, and JCJC INS2I 2017 ``QFCCQI'' project.

\bibliographystyle{WileyNJD-AMA}
\bibliography{mybibfile}

\section{Appendix}\label{appendix}

\begin{theorem}\label{theorem2}
Let $\{Y_{i,t},i=1,2\}$ be two homodyne detection measurements for a two-level quantum system. With single-photon input field, the quantum filter for the conditional expectation in the Heisenberg picture is given by
\begin{scriptsize}
\begin{eqnarray}\label{qqhei}\begin{aligned}
d\pi^{11}_t(X)=&\left\{\pi^{11}_t(\mathcal{L}_GX)+\pi^{01}_t(S^\dag[X,L])\xi^\ast(t)+\pi^{10}_t([L^\dag,X]S)\xi(t)+\pi^{00}_t(S^{\dag}XS-X)|\xi(t)|^2\right\}dt\\
&+\Big[s_{11}\pi^{11}_t(XL)+s_{11}^\ast\pi^{11}_t(L^{\dag}X)+s_{11}\pi^{10}_t(XS)\xi(t)
+s_{11}^\ast\pi^{01}_t(S^{\dag}X)\xi^\ast(t)-\pi^{11}_t(X)K_1(t)\Big]dW_1(t)\\
&+\Big[s_{21}\pi^{11}_t(XL)+s_{21}^\ast\pi^{11}_t(L^{\dag}X)+s_{21}\pi^{10}_t(XS)\xi(t)
+s_{21}^\ast\pi^{01}_t(S^{\dag}X)\xi^\ast(t)-\pi^{11}_t(X)K_2(t)\Big]dW_2(t),\\
d\pi^{10}_t(X)=&\left\{\pi^{10}_t(\mathcal{L}_GX)+\pi^{00}_t(S^\dag[X,L])\xi^\ast(t)\right\}dt\\
&+\left[s_{11}\pi^{10}_t(XL)+s_{11}^\ast\pi^{10}_t(L^{\dag}X)+s_{11}^\ast\pi^{00}_t(S^{\dag}X)\xi^\ast(t)-\pi^{10}_t(X)K_1(t)\right]dW_1(t)\\
&+\left[s_{21}\pi^{10}_t(XL)+s_{21}^\ast\pi^{10}_t(L^{\dag}X)+s_{21}^\ast\pi^{00}_t(S^{\dag}X)\xi^\ast(t)-\pi^{10}_t(X)K_2(t)\right]dW_2(t),\\
d\pi^{00}_t(X)=&\pi^{00}_t(\mathcal{L}_GX)dt+\left[s_{11}\pi^{00}_t(XL)+s_{11}^\ast\pi^{00}_t(L^{\dag}X)-\pi^{00}_t(X)K_1(t)\right]dW_1(t)\\
&+\left[s_{21}\pi^{00}_t(XL)+s_{21}^\ast\pi^{00}_t(L^{\dag}X)-\pi^{00}_t(X)K_2(t)\right]dW_2(t),
\end{aligned}\end{eqnarray}
\end{scriptsize}
where
\begin{eqnarray}\begin{aligned}
K_1(t)&=s_{11}^\ast\pi_t^{11}(L^\dag)+s_{11}\pi_t^{11}(L)+s_{11}^\ast\pi_t^{01}(S^\dag)\xi^\ast(t)+s_{11}\pi_t^{10}(S)\xi(t),\\
K_2(t)&=s_{21}^\ast\pi_t^{11}(L^\dag)+s_{21}\pi_t^{11}(L)+s_{21}^\ast\pi_t^{01}(S^\dag)\xi^\ast(t)+s_{21}\pi_t^{10}(S)\xi(t),
\end{aligned}\end{eqnarray}
and $\pi^{01}_t(X)=(\pi^{10}_t(X^\dagger))^\dagger$, the Wiener processes $W_1(t)$ and $W_2(t)$ are given by
\begin{equation}
dW_1(t)=dY_{1,t}-K_1(t)dt,~~dW_2(t)=dY_{2,t}-K_2(t)dt,
\end{equation}
respectively. The initial conditions are $\pi_0^{11}(X)=\pi_0^{00}(X)=\langle\eta,X\eta\rangle$, $\pi_0^{10}(X)=\pi_0^{01}(X)=0$.
\end{theorem}

By the filter equations \eqref{qqhei} and $\pi^{jk}_t(X)=\mathrm{Tr}[(\rho^{jk}(t))^{\dag}X]$, we can also present the quantum filter in the Schr\"{o}dinger picture.

\begin{corollary}\label{corollary2}
With two homodyne detection measurements, the quantum filter for a two-level quantum system driven by single-photon input field in the Schr\"{o}dinger picture is given by
\begin{scriptsize}
\begin{eqnarray}\label{qqsch}\begin{aligned}
d\rho^{11}(t)=&\left\{\mathcal{L}^\star_G\rho^{11}(t)+[S\rho^{01}(t),L^\dag]\xi(t)+[L,\rho^{10}(t)S^\dag]\xi^\ast(t)+[S\rho^{00}(t)S^\dag-\rho^{00}(t)]|\xi(t)|^2\right\}dt\\
&+\left[s_{11}^\ast\rho^{11}(t)L^\dag+s_{11}L\rho^{11}(t)+s_{11}^\ast\rho^{10}(t)S^\dag\xi^\ast(t)+s_{11}S\rho^{01}(t)\xi(t)-K_1(t)\rho^{11}(t)\right]dW_1(t)\\
&+\left[s_{21}^\ast\rho^{11}(t)L^\dag+s_{21}L\rho^{11}(t)+s_{21}^\ast\rho^{10}(t)S^\dag\xi^\ast(t)+s_{21}S\rho^{01}(t)\xi(t)-K_2(t)\rho^{11}(t)\right]dW_2(t),\\
d\rho^{10}(t)=&\left\{\mathcal{L}^\star_G\rho^{10}(t)+[S\rho^{00}(t),L^\dag]\xi(t)\right\}dt\\
&+\left[s_{11}^\ast\rho^{10}(t)L^\dag+s_{11}L\rho^{10}(t)+s_{11}S\rho^{00}(t)\xi(t)-K_1(t)\rho^{10}(t)\right]dW_1(t)\\
&+\left[s_{21}^\ast\rho^{10}(t)L^\dag+s_{21}L\rho^{10}(t)+s_{21}S\rho^{00}(t)\xi(t)-K_2(t)\rho^{10}(t)\right]dW_2(t),\\
d\rho^{00}(t)=&\mathcal{L}^\star_G\rho^{00}(t)dt+\left[s_{11}^\ast\rho^{00}(t)L^\dag+s_{11}L\rho^{00}(t)-K_1(t)\rho^{00}(t)\right]dW_1(t)\\
&+\left[s_{21}^\ast\rho^{00}(t)L^\dag+s_{21}L\rho^{00}(t)-K_2(t)\rho^{00}(t)\right]dW_2(t),
\end{aligned}\end{eqnarray}
\end{scriptsize}
where
\begin{eqnarray}\begin{aligned}
K_1(t)=&s_{11}\mathrm{Tr}[L\rho^{11}(t)]+s_{11}^\ast\mathrm{Tr}[L^\dag\rho^{11}(t)]+s_{11}\mathrm{Tr}[S\rho^{01}(t)]\xi(t)+s_{11}^\ast\mathrm{Tr}[S^\dag\rho^{10}(t)]\xi^\ast(t),\\
K_2(t)=&s_{21}\mathrm{Tr}[L\rho^{11}(t)]+s_{21}^\ast\mathrm{Tr}[L^\dag\rho^{11}(t)]+s_{21}\mathrm{Tr}[S\rho^{01}(t)]\xi(t)+s_{21}^\ast\mathrm{Tr}[S^\dag\rho^{10}(t)]\xi^\ast(t),
\end{aligned}\end{eqnarray}
and $\rho^{01}(t)=(\rho^{10}(t))^\dagger$, the initial conditions are $\rho^{11}(0)=\rho^{00}(0)=|\eta\rangle\langle\eta|$, $\rho^{10}(0)=\rho^{01}(0)=0$.
\end{corollary}

\end{document}